\documentclass[journal]{IEEEtran}
\usepackage{wrapfig} 
\usepackage{mathtools}
\usepackage{graphicx,dblfloatfix}
\usepackage{epstopdf}
\usepackage{lettrine}
\usepackage{cite}
\usepackage{url}
\newcommand{\beq}{\begin{equation}}
\newcommand{\eeq}{\end{equation}}
\newcommand{\beqn}{\begin{eqnarray}}
\newcommand{\eeqn}{\end{eqnarray}}
\begin{document}

\title{Performance Analysis of Multiuser FSO/RF Network Under Non-Equal Priority with $P$-Persistence Protocol}
\author{Tamer~Rakia,
Fayez Gebali,~\IEEEmembership{Senior Member,~IEEE},
Hong-Chuan~Yang,~\IEEEmembership{Senior Member,~IEEE}, and
 Mohamed-Slim Alouini,~\IEEEmembership{Fellow,~IEEE}}
\maketitle
\begin{abstract}
This paper presents and analyzes a novel multiuser network based on hybrid free-space optical (FSO)/radio-frequency (RF) transmission system, where every user is serviced by a primary FSO link.
When more than one FSO link fail, the central node services these corresponding users by using a common backup RF link according to a non-equal priority with $p$ persistence servicing protocol. 
A novel discrete-time Markov chain model is developed for the proposed network, where different transmission rates over both RF and FSO links are assumed.  We investigate the throughput from central node to the user, the average size of the transmit buffer allocated for every user, the frame queuing delay in the transmit buffer, the efficiency of the queuing system, the frame loss probability, and the RF link utilization. The network's users are arranged in an ascending order. As the order of the failed user increases, the corresponding performance criteria gets worse due to the non-equal priority servicing protocol. Numerical examples show that  transmitting a data frame with probability $p$ when using the common backup RF link, achieves considerable performance improvement that approaches the performance of the multiuser FSO/RF network when using equal priority protocol to serve all the remote users. 
\end{abstract}

\section{Introduction}
\lettrine{R}                        
ECENTLY free-space optical (FSO) technology has gained an increasing interest in data transmission \cite {Kn:FSO0} --\cite {kn:FSO3} owing to their unique features including: 1) higher data rate, compared to radio frequency (RF) technology, 2) high transmission security, 3) large unregulated spectrum, and 4) faster and cheaper deployments, compared to fiber optics. 
Most of the current literature on FSO transmission systems are mainly limited to connecting two remote locations through a direct point-to-point FSO  transmission link. Meanwhile, FSO technology can also be used effectively in multiuser scenarios \cite{kn:multiuser0} --\cite{kn:multiuser2}, where FSO links are used for data transmission from a central location to multiple remote users. In our previous work \cite{kn:multiuser0}, a cross layer analysis of a point-to-multipoint hybrid FSO/RF network was performed where same data rates on both FSO and backup RF links was assumed. Several performance parameters of the network were investigated. 
In \cite{kn:multiuser1} and \cite{kn:multiuser2}, the performance of multiuser FSO systems were analyzed, where in \cite{kn:multiuser1} authors had investigated the average bit error rate and the capacity of the systems with $Nth$ best user selection. While in \cite{kn:multiuser2}, the outage capacity and the throughput of the systems were studied.

Despite the desirable nature of the FSO systems, FSO links are greatly affected by fading due to atmospheric turbulences and pointing errors \cite{kn:Farid}, \cite{kn:Imran}. Turbulence-induced fading, known as scintillation, causes irradiance fluctuations in the received optical signal as a result of variations in the atmospheric refractive index. Dynamic wind loads and weak earthquakes can cause vibrations of the transmitted optical beam, which also causes random irradiance fluctuations in the received optical signal. Moreover, the optical power is attenuated as the distance between the transmitter and the receiver increases due to a constant loss related to the weather condition. 
Integrating the FSO link with a millimeter wavelength (MMW) RF link, to form what is known by hybrid FSO/RF data transmission link, improved the FSO link's reliability. This is owed to the fact that both FSO and MMW RF links are affected quite differently by atmospheric and weather effects. FSO links suffer from extremely high attenuation in the presence of fog but are less affected by rain. In contrast, fog has practically no effect on MMW RF links but rain significantly increases link attenuation. Similarly, while atmospheric turbulence caused by variations in the refractive index is the main cause of small--scale fading in FSO links \cite{kn:gamma}, RF links are impaired by fading due to multipath propagation \cite{kn:Guo}. 
Besides the high data rates, MMW RF links offer similar advantages of FSO links such as deployment flexibility, license free operation, and inherent security due to high link attenuation. 
The complementary nature of FSO and MMW RF links has motivated various hybrid FSO/RF data transmission systems which include diversity combining hybrid FSO/RF systems \cite{kn:ref10}, switch-over hybrid FSO/RF systems \cite{kn:ref110} and hybrid FSO/RF systems with adaptive combining \cite{kn:tamer}. Hybrid FSO/RF data transmission systems had shown great outage performance and high data link reliability in all weather conditions, while maintaining the high transmission data rate. 

In this paper, we extend our work in \cite{kn:multiuser0}, where we consider a multiuser hybrid FSO/RF network with multiple remote users are connected to one single central location.
Each remote user in the network is connected to the central node via a separable primary FSO link. 
A common backup RF link is used by the central node for data transmission to any remote user in case of the failure of its corresponding FSO link, where we adopt the simple switch-over hybrid FSO/RF transmission approach. 
A direct application of the proposed multiuser hybrid FSO/RF network is wireless Internet service provider (WISP) networks or the WiMAX networks \cite{Kn:PTMP1}. 
In a WISP network, subscribers are connected at the edge of the network using a client device typically mounted on the roof of their houses. The central base station is mounted on a high building where it has line of sight with the client devices.
At any given time slot, there may be more than one remote user that has failed FSO link. The central node must adopt some strategy for selecting one user among these users to be serviced by the common backup RF link. To do this, there are many options that include:
\begin{enumerate}
\item{Select remote user with highest number of packets in buffer.}
\item{Select remote users at random, which is similar to the back-off counter in WiFi IEEE 802.16.}
\item{Assign a probability $p$ to each remote user to determine the odds that the user is to compete at this time slot or not, which is similar to carrier-sense multiple access with collision detection (CSMA/CD) or carrier-sense multiple access with collision avoidance (CSMA/CA) schemes.}
\item{Adopt a fixed priority scheme where each remote user has a unique priority value. Remote user with highest priority among the set of users with failed FSO links is selected.}
\item{Group remote users in different classes of service, then select a user at random from the highest priority class.}
\item{Adopt the proportional fair (PF) scheduling scheme similar to long term evolution (LTE) networks.}
\end{enumerate}
In this proposed infrastructure network, the central node adopts a hybrid priority scheme, which is a combination of options 3 and 4. Priority services can be used to provide differentiated Quality-of-Service (QoS) in packet-based networks. Example for different services needing different QoS can include general communications, information retrieval, web services, and streaming services. Such networks offering this service model will charge users based on the priority of their traffic. 
Thus, considering the differentiated QoS, users are arranged according to their servicing priorities. The central node serves the different nodes according to the predetermined priority scheme, but with $p$-persistence probability. 
We will call this servicing protocol ``non-equal priority with $p$-persistence servicing protocol". Controlling the value of $p$ will control the whole network's performance.
 It is worth to mention here that, the proposed hybrid priority servicing scheme is different from the LTE PF scheduling scheme. The proposed servicing scheme provides a compromise to give better performance to high-priority users while allowing lower priority users a chance to receive their intended data. In the mean time, the total throughput can be maximized by optimizing $p$, as will be discussed in coming sections. Meanwhile, PF scheduler tries to maximize the total network throughput, while  providing all users a minimal level of service at the same time. PF scheduling scheme achieves a trade-off between fairness of resources allocation to all users and network throughput.

In this paper, we investigate such a multiuser network under the proposed servicing protocol through investigating several performance criteria. These criteria include throughput from central node to the remote users, the average size of the transmit buffer assigned for every remote user in the network, the frame queuing delay in the transmit buffer, the efficiency of the queuing system, the frame loss probability, and the RF link utilization. To be able to investigate such a network under this hybrid priority servicing scheme, we assume identical remote users, while differ according to their required QoS.

The major contributions in this paper are: 
\begin{itemize}
\item{A novel multiuser hybrid FSO/RF network is proposed where different transmission rates over FSO and RF links are considered and non-equal priority with $p$-persistence servicing protocol is proposed.}
\item{A novel cross layer Markov chain model of this proposed multiuser network with different data rates is developed.}
\item{The main parameters affecting the performance of the proposed multiuser hybrid FSO/RF network are identified, and several performance metrics are studied.}
\end{itemize}

The remainder of the paper is organized as follows. In section II and III, we introduce the multiuser hybrid FSO/RF network model and central node-to-remote node channel model, respectively. In section IV, we introduce the discrete-time Markov chain model for a remote node in the multiuser network. The non-equal priority with $p$-persistence servicing protocol is presented in section V. Different performance metrics for the multiuser network are studied in section VI. Finally, section VII presents some numerical examples to investigate the performance of the proposed multiuser network, followed by the conclusion in section VIII. 

\section{Multiuser Hybrid FSO/RF Network Modeling}
\begin{figure}[ht]
\centering
\includegraphics [height=5.0cm, width=5.0cm]{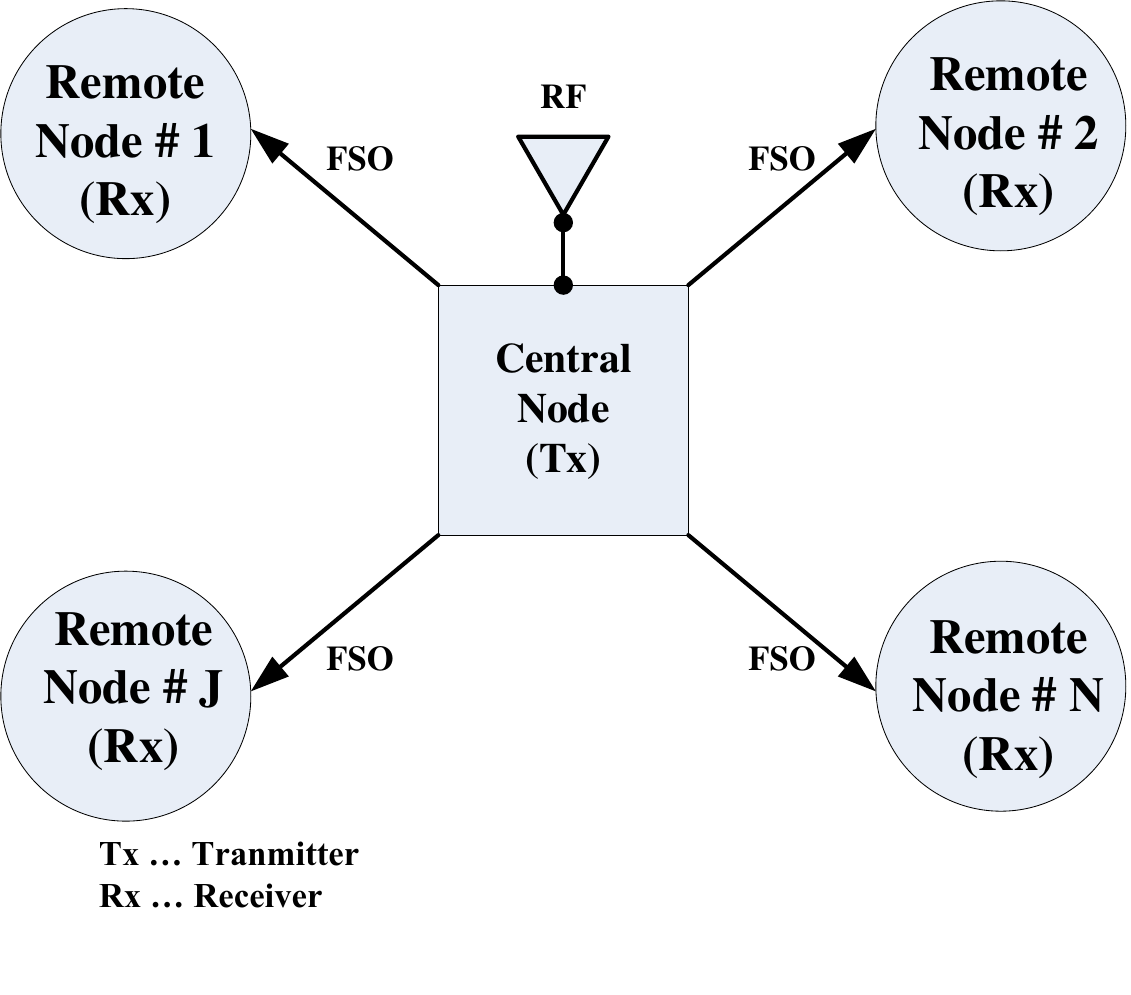} 
\caption{General block diagram of a multiuser hybrid FSO/RF network.}
\label{network_block}
\end{figure}

The general block diagram of a multiuser hybrid FSO/RF network is shown in Fig. \ref{network_block}.
This network consists of a central node and $N$ remote users (remote nodes). 
The central node sends data to each one of the remote users through a separate primary FSO link.
The central node is assigned a certain RF channel, which is used as a backup link for data transmission to any remote user in case of the failure of its corresponding FSO link. 
In this work, we propose and study a non-equal priority with $p$-persistence protocol for servicing the remote users by using the common backup RF link in case of the failure of their corresponding FSO links. The remote users are arranged in an ascending order according to their servicing priorities to be $1, 2, ..., N$, where the lowest order user will have the highest priority. In this case, when more than one FSO link fail, the central node will use the common backup RF link to send data frame with probability $p$ to the remote user with the highest priority among the failed group.

Throughout the paper, we will adopt the following assumptions:
\begin{enumerate}
\item{All FSO links follow the same statistical fading distribution.}
\item{The FSO channel is assumed to be constant during the transmission of one data frame.}
\item{The RF channel is assumed to be constant during the transmission of one data frame.}
\item{Simultaneous data transmission over FSO and RF links is not allowed.}
\item{Non-preemptive RF transmission is assumed.}
\item{Non-saturated traffic condition is assumed. Specifically, the central node may or may not have data for transmission to all $N$ remote users over each time step.} 
\item{The data frames are transmitted over the FSO and RF links at different rates.}
\item{Once the data frame is being transmitted over the backup RF channel, it is assumed that the RF channel will remain healthy throughout the whole frame transmission.}
\item{No error control policy is implemented between the remote user and the central node.}
\item{The central node receives 
the states of the $N$ FSO channels and the common backup RF channel based on feedback from the $N$ remote users\footnote{We assume that there is an RF feedback link between each remote user and the central node. The central node transmits pilots over the common backup RF channel, which can be received by the remote users. These pilots allow the estimation of the CSI of the common RF channel at the remote users ($N$ CSIs), which will be sent back to the central node over the RF feedback link. The CSI of the $N$ FSO links can also be estimated and sent back to central node over the RF feedback links. The RF feedback link can be a common low rate RF channel that can be used by all $N$ remote users, where each user sends its CSI along with an identifier differentiating users. As a commonly used assumption, the RF feedback link is assumed to be error free.}. Since both FSO and MMW RF channels experience slow-fading \cite{kn:Farid}, \cite{kn:Nick}, we can assume that instantaneous CSI is always available at the transmitters.} 
\end{enumerate}
\section{Central Node - Remote Node Channel Modelling}
The hybrid FSO/RF system used for data transmission from the central node to any remote node is composed of coherent/heterodyne FSO and RF data transmission subsystems. In this hybrid system, coded digital baseband signal, created by the signal source, is converted to an analog electrical signal through a square $M$--ary quadrature amplitude modulation (QAM) modulator \cite{kn:MQAM_FSO},\cite{kn:MQAM_RF}. 
The QAM signal will be transmitted using either the FSO link or the RF link.
At the FSO transmitter, QAM electrical signal modulates an optical carrier, produced by an optical frequency local oscillator (LO) to generate the optical signal. At the FSO receiver, the received optical signal undergoes heterodyne detection process. With heterodyne detection, the received information bearing optical signal is combined with a reference signal from an optical LO using a beam splitter\footnote{We assume using phase-locked loop to compensate for phase noise in the received optical signal \cite{kn:coherent}.}. When the combined optical signal falls over a photo-detector, an instantaneous photo-current is produced. After filtering unnecessary components, the photo-current become the input of the electrical demodulator to retrieve the information.
At the RF transmitter, the frequency of the QAM electrical signal is up-converted using 60 GHz RF carrier, generated by an RF LO.
We adopt a switch-over hybrid FSO/RF transmission scheme to transmit the data frames from the central node to any remote node \cite{kn:multiuser0}.
We define $\gamma_T$ to be the minimum  received SNR of either the FSO link or the common RF link to maintain the minimum bit-error-rate of $\mathrm{BER}_0$.
Considering the used $M$-QAM electrical modulation scheme and by using \cite[Eq. (9.7)]{kn:book_goldsmith}, the switching threshold $\gamma_T$ can be calculated as $\gamma_T = (M-1) \left[-\frac{2}{3}\ln(5~ \mathrm{BER}_0)\right],  \ M\ge4$.

\subsection{Modelling the FSO Link}
The FSO link is assumed to be affected by a  Gamma - Gamma atmospheric turbulence-induced fading and a Gaussian pointing-error induced fading.
We define $a$ to be the probability that a certain FSO link is in poor quality and can not be used for data transmission from the central node to the corresponding remote node. $a$ is given by $a = {P_r}[\gamma_{FSO} < \gamma_T] =  F_{\gamma_{FSO}}(\gamma_T)$,
where $F_{\gamma_{FSO}}(.)$ is the cumulative distribution function (CDF) of the instantaneous SNR per symbol of the FSO receiver $\gamma_{FSO}$ defined as \cite{kn:multiuser0}:
\begin{equation}
\label{CDF SNR FSO}
\begin{split}
F_{\gamma_{FSO}}(\gamma_ {FSO})=&\frac{\xi^2}{\Gamma(\alpha)\Gamma(\beta)} G_{2,4}^{3,1}\left[\frac{\xi^2\alpha\beta \gamma_{FSO}}{ (\xi^2 + 1)\bar\gamma_{{FSO}}} \mid_{\xi^2,~\alpha,~\beta,~0}^{1,~\xi^2 + 1}\right].
\end{split}
\end{equation}
In (\ref{CDF SNR FSO}), $\xi$ is the ratio between the equivalent beam radius and the pointing error (jitter) standard deviation $\sigma_s$  given by $\xi = \omega_{eq}/2\sigma_s$ \cite{kn:Farid}. Here, $\omega_{eq}^2=\omega_z^2 \sqrt{\pi} \text{erf}(\nu)/2 \nu \exp(-\nu^2)$, where $\text{erf}(.)$ is the error function and $\omega_z$ is the optical beam radius at distance $z$ from the transmitter aperture and $\nu=\sqrt{\pi} D/2\sqrt{2} \omega_z$ with $D$ is the photodetector diameter. $\omega_z$ is given by $\omega_z = \theta_0 z$, where $\theta_0$ is the transmit divergence at $1/e^2$. $\Gamma(.)$ and $G[.]$ are respectively, the standard Gamma function and the Meijer G-function as defined in \cite [Eq. (9.301)]{kn:book_it02} with $\alpha$ and $\beta$ are the scintillation parameters. The parameters $\alpha$ and $\beta$ are related to the refractive index structure parameter $C_n^2$, where the atmospheric turbulence can be modelled from weak to strong turbulence regimes \cite{kn:book_it06}. 
Assuming spherical optical wave propagation, expressions for calculating $\alpha$ and $\beta$ in (\ref{CDF SNR FSO}) are given by \cite{kn:book_it06}: 
\begin{equation}
\label{alpha2}
\alpha=\left[\exp\left(\frac{0.49\chi^2}{(1 + 0.18d^2 + 0.56\chi^\frac{12}{5})^\frac{7}{6}}\right)-1\right]^{-1}
\end{equation}
\begin{equation}
\label{beta2}
\beta=\left[\exp\left(\frac{0.51\chi^2 (1+0.69\chi^\frac{12}{5})^{-\frac{5}{6}}}{(1+ 0.9d^2 + 0.62 d^2 \chi^\frac{12}{5})^\frac{5}{6}}\right)-1\right]^{-1},
\end{equation}
where $\chi^2 = 0.5 C_n^2 k^{7/6} z^{11/6}$ is the Rytov variance, $d=(kD^2/4z)^{1/2}$, and $k = 2\pi/\lambda_{FSO}$ is the optical wave number with $\lambda_{FSO}$ is the optical wavelength.
Obviously, the scintillation parameters $\alpha$ and $\beta$ and the pointing-error parameter $\xi$ may differ from one remote node to another. However, without loss of generality and to make the analysis easy and tractable, we assume that the parameters $\alpha$, $\beta$, and $\xi$ are the same for all remote nodes\footnote{We assume all remote nodes of the network are at equidistant from the central node.}.
$\bar\gamma_{FSO}$ is the average SNR, defined as $\bar\gamma_{FSO}={2E_{avg} \eta^2 P_{LO} Pow_{FSO} G_{FSO}}/{\sigma_{FSO}^2}$ \cite{kn:multiuser0}, where $E_{avg}$, $\eta$, $P_{LO}$ , $Pow_{FSO}$, $G_{FSO}$, and $\sigma_{FSO}^2$ are respectively, the average QAM symbol energy, photodetector responsivity, LO power, average transmitted optical power\footnote{The transmitted optical power is constrained by the skin and eye safety requirements according to the used wavelength \cite{kn:book_Hranilovic}.}, attenuation factor, and variance of shot noise which is modelled as additive white Gaussian noise (AWGN). The attenuation factor $G_{FSO}$ (in dB) is determined by the Beers-Lambert law as $G_{FSO} = \alpha_{FSO} z$ \cite{kn:book_Beers}, where $\alpha_{FSO}$ denotes the weather attenuation coefficient (in dB/Km) and $z$ is the link distance.

\subsection{Modelling the RF Link} 
The fading gain over the RF channel is assumed to follow Nakagami-$m$ distribution \cite{kn:tamer}, which represents a wide variety of realistic line-of-sight (LOS), (with high values of the parameter $m$ \cite{kn:book_alouini}), and non LOS fading channels encountered in practice \cite{kn:nak-m}.
As we are using the same type of digital modulation on both FSO and RF links, we will use the same threshold $\gamma_T$ to measure the quality of the RF link from the central node to the remote node. We define $b$ to be the probability that the RF link is in poor quality and can't be used for data transmission by the central node, given by $b = {P_r}[\gamma_{RF} < \gamma_T] =  F_{\gamma_{RF}}(\gamma_T)$,
where $F_{\gamma_{RF}}(.)$ is the CDF of the instantaneous SNR per symbol of the RF receiver $\gamma_{RF}$ defined as \cite{kn:multiuser0}:
\begin{equation}
\label{CDF SNR RF}
F_{\gamma_{RF}}(\gamma_{RF})=\frac{1}{\Gamma(m)}\gamma\left(m,\frac{m\gamma_{RF}}{\bar\gamma_{RF}}\right). 
\end{equation}
In (\ref{CDF SNR RF}), $\gamma(\cdot,\cdot)$ is the lower incomplete Gamma function defined in \cite [Eq. (8.350.1)]{kn:book_it02}, $\bar\gamma_{RF}$ is the average SNR of the RF channel, given by $\bar\gamma_{RF} ={E_{avg}P_{RF}G_{RF}}/{\sigma_{RF}^2}$ \cite{kn:multiuser0}, where $P_{RF}$, $\sigma_{RF}^2$, and $G_{RF}$ are respectively, transmitted RF power\footnote{The transmitted RF power is constrained by the allowable interference levels that may be introduced to the systems working at the same RF band and the radiation constraint.}, noise variance, assuming zero-mean circularly symmetric AWGN, and average power gain of the RF channel. The variance of the noise in the RF channel is given by  $\sigma_{RF}^2 = W N_0 N_F$ \cite{kn:ref10}, where $W$ is the RF bandwidth, $N_0$ is the noise power spectral density and $N_F$ is the noise figure of the RF receiver.
\section{Discrete-Time Markov Chain Model for $J$th Remote Node}
\begin{figure}[t]
\centering
\includegraphics [height=8.0cm, width=9.0cm]{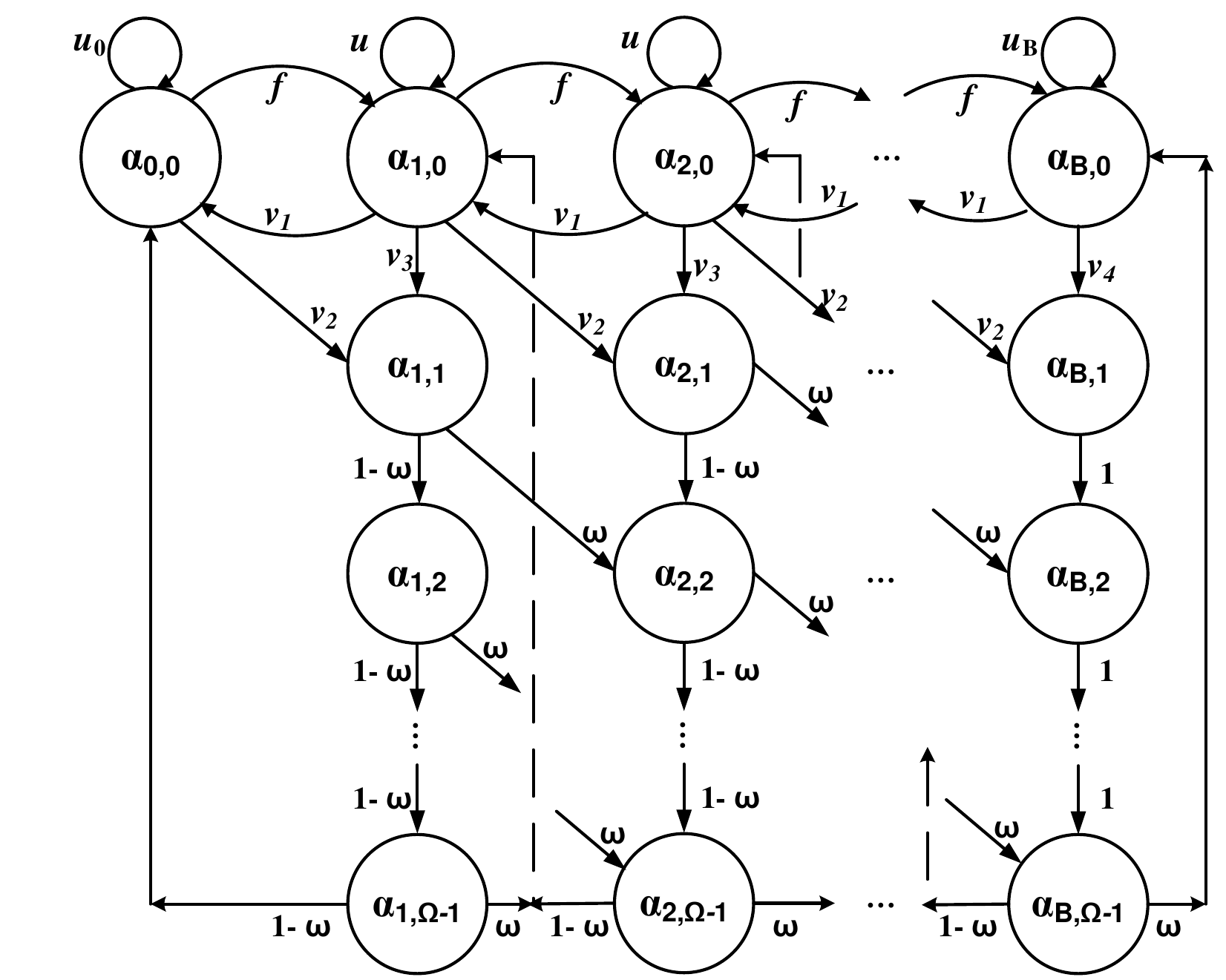} 
\caption{The state transition diagram for the transmit buffer of the $J$th remote node.}
\label{state_diagram}
\end{figure}

Since all the $N$ remote nodes are identical, we can study the proposed multiuser network performance by focusing on one remote node, which we call the $J$th remote node. We will use the term ``other nodes" to refer to the remaining ($N$-1) remote nodes. 
The central node assigns a first-in-first-out (FIFO) transmit buffer of size $B$ frames for every remote node $J \in [1, 2, ... ,N]$. This buffer will hold the data frames to be transmitted from the central node to the corresponding remote node. The data frames arrive at the transmit buffer at rate $R_{in}$ frames/second. This frame arrival rate $R_{in}$ is related to the input data rate $r$ in bits/second by $R_{in} = N_{s}r/ \log_2 M$, where $N_s$ is the number of symbols in each frame and $M$ is the order of the $M$-QAM scheme used. 
The frame arrival rate $R_{in}$ is assumed to be the same for all the $N$ transmit buffers. 
We define $R_{out}$ as the frame departure rate given by:
\begin{equation*}
R_{out} = 
\begin{cases}
R_{FSO},&\text{If data frame is transmitted over FSO link}\\
R_{RF},&\text{If data frame is transmitted over RF link},
\end{cases}
\end{equation*}
where $R_{FSO}$ and $R_{RF}$ are the frame transmission rates over FSO and RF links, respectively. $R_{FSO}$ and $R_{RF}$ are related by $R_{FSO} =  \Omega  R_{RF}, \ \Omega \ge 1$.
It is worth to mention that $\Omega$ will be restricted to integer values. This point will be clear later from the state diagram shown in Fig. \ref{state_diagram}.
To prevent transmit buffer overflow, we must ensure that $R_{out}$ is greater than or equal to $R_{in}$. 
Here, the frame arrival rate $R_{in}$ changes over time with maximum value equal to $R_{out}$. 

The number of frames stored in the transmit buffer represents the state of the buffer. 
The future state of the transmit buffer depends only on its current state and the change from one state to another will occur at discrete time values corresponding to frame arrival and departure events. Thus, we can use the discrete-time Markov chain 
to model the states of the transmit buffer assigned for any remote node. 
The time step of the discrete-time Markov chain, denoted by $T$, is chosen to be the inverse of the maximum frame transmission rate, as $T = {1}/{\text{max}(R_{in}, R_{out})} = {1}/{R_{FSO}} = {1}/{\Omega  R_{RF}}$ \cite{kn:book_Fayez}.
It is worth to mention here that one data frame will need one time step to be transmitted over the FSO link and $\Omega$ time steps to be transmitted over the RF link.

Based on our choice for the time step $T$, the resulting Markov chain is a single-arrival, single-departure queue. Here, we assume that when a data frame arrives, it can be serviced and left the transmit buffer at the same time step using the corresponding FSO link or can go to the next state using the RF link with a lower data rate. The corresponding state transition diagram of the transmit buffer of the $J$th remote node is shown in Fig. \ref{state_diagram}. We define $\alpha_{i,j}$ for $0\leq i \leq B$ and $0\leq j \leq \Omega-1$ to be the transmit buffer states for the $J$th remote node. Here, $i$ represents the number of data frames stored in the buffer and $j$ represents the number of time steps elapsed since the beginning of one data frame transmission over the RF link. 

In Fig. \ref{state_diagram}, $u_0$, $f$, $u$, $u_B$, $v_1$, $v_2$, $v_3$, and $v_4$ are the states transition probabilities defined as:
\begin{equation}
\begin{split}
u_0 &= 1 - \omega + \omega P_{FSO},  f = \omega [1 - (P_{FSO}+ P^{(J)}_{RF})],\\
u &= (1 - \omega)[1 - (P_{FSO}+ P^{(J)}_{RF})] + \omega P_{FSO},\\
u_B &= 1 - (P_{FSO}+ P^{(J)}_{RF}) + \omega P_{FSO}, v_1 = (1-\omega) P_{FSO},\\
v_2 &= \omega P^{(J)}_{RF}, v_3 = (1-\omega) P^{(J)}_{RF}, v_4 = P^{(J)}_{RF},
\end{split}
\label{stp}
\end{equation}
where $P_{FSO}$ is the probability of using the FSO link by the central node to transmit data frame to the corresponding $J$th remote node, given by $P_{FSO} = 1 - a$.
In (\ref{stp}), $P^{(J)}_{RF}$ is the probability of using the common backup RF link by the central node to transmit data frame to the $J$th remote node when its corresponding FSO link fails. The definition of $P^{(J)}_{RF}$ will depend on the type of the protocol of assigning the RF link to the $J$th remote node, either equal priority or non-equal priority with $p$-persistence protocol. 
In (\ref{stp}), $\omega$ is the data frame arrival probability defined as the probability that a new data frame arrives at the transmit buffer assigned for the $J$th remote node within the time step $T$. Or in other words, $\omega$ is the probability that a time step $T$ will have a new data frame.

The corresponding state transition matrix $\textbf{P}^{(J)}$ for the state diagram in Fig. \ref{state_diagram}  is given by: 
\setlength{\abovedisplayskip}{5pt}
\begin{equation}
\label{Eq_P}
\textbf{P}^{(J)}= \left[ \begin{array}{ccccccc}
A_0 & C_0 &\phi_1& ... &\phi_1 & \phi_1 &\phi_1\\
E_0            & A        & C & ... & \phi_2 & \phi_2 & \phi_2\\
\phi_1^T   & E         & A & ... & \phi_2 & \phi_2 & \phi_2\\
\phi_1^T   &\phi_2  & E & ...  & \phi_2 & \phi_2 & \phi_2\\
\vdots & \vdots & \vdots & \ddots & \vdots & \vdots& \vdots\\
\phi_1^T     & \phi_2      & \phi_2 & ... & C & \phi_2 &\phi_2\\
\phi_1^T     & \phi_2      & \phi_2 & ... & A& C &\phi_2\\
\phi_1^T     & \phi_2      & \phi_2 & ... & E & A & C\\
\phi_1^T     & \phi_2      & \phi_2 & ... & \phi_2 & E & A_B\\
\end{array} \right]_{\Omega B+1 \times \Omega B+1},
\end{equation}
where $A_0$, $C_0$, $E_0$, $A$, $C$, $E$, $A$, $A_B$, $\phi_1$, and $\phi_2$ are sub-matrices defined as:\\
$
A_0\ = [u_0]_{1\times1}, C_0\ = \left[ \begin{array}{ccccc}
v_1 & 0 & ... &0 &1-\omega
\end{array} \right]_{1\times \Omega},\\
C_0\ = \left[ \begin{array}{ccccc}
v_1 & 0 & ... &0 &1-\omega
\end{array} \right]_{1\times \Omega},\\
E_0\ = \left[ \begin{array}{ccccc}
f& v_2 & 0 & ... &0  \quad \ \ 
\end{array} \right]^T_{1\times \Omega},\\
A\ \ = \left[ \begin{array}{ccccccc}
u &  0 & 0 &...&0 & \omega \\
v_3 & 0 & 0 &... & 0 & 0\\
0     & 1-\omega & 0& ...& 0 & 0\\
0     & 0 & 1-\omega& ...& 0 & 0\\
\vdots & \vdots & \vdots & \ddots & \vdots  & \vdots \\
0     & 0 & 0& ...& 1-\omega & 0\\
\end{array} \right]_{\Omega \times \Omega},\\
C\ \ = \left[ \begin{array}{ccccccc}
v_1 &  0 & 0 &...&0 &1- \omega \\
0 & 0 & 0 &... & 0 & 0\\
\vdots & \vdots & \vdots & \ddots & \vdots  & \vdots \\
0& 0 & 0& ...& 0 & 0\\
\end{array} \right]_{\Omega \times \Omega},\\
E\ \ = \left[ \begin{array}{ccccccc}
f &  0 & 0 &...&0 & 0 \\
v_2 & 0 & 0 &... & 0 & 0\\
0     & \omega & 0& ...& 0 & 0\\
0     & 0 & \omega& ...& 0 & 0\\
\vdots & \vdots & \vdots & \ddots & \vdots  & \vdots \\
0     & 0 & 0& ...& \omega & 0\\
\end{array} \right]_{\Omega \times \Omega},\\
A_B= \left[ \begin{array}{ccccccc}
u_B &  0 & 0 &...&0 & \omega \\
v_4 & 0 & 0 &... & 0 & 0\\
0     & 1 & 0& ...& 0 & 0\\
0     & 0 & 1& ...& 0 & 0\\
\vdots & \vdots & \vdots & \ddots & \vdots  & \vdots \\
0  & 0 & 0& ...& 1 & 0\\
\end{array} \right]_{\Omega \times \Omega},\\
\phi_1\ \ = \big[\begin{array}{cccccc}
0& 0& 0 & ... &0&0
\end{array} \big]_{1\times \Omega},\\
\phi_2\ \ = \left[ \begin{array}{ccccccc}
0 &  0 & 0 &...&0 & 0 \\
0 & 0 & 0& ...& 0 & 0\\
\vdots & \vdots & \vdots & \ddots & \vdots  & \vdots \\
0 & 0 & 0& ...& 0 & 0\\
\end{array} \right]_{\Omega \times \Omega}.\\
$

The steady state distribution vector $\textbf{s}^{(J)}$ corresponding to the state diagram in Fig. \ref{state_diagram} and the transition matrix (\ref{Eq_P}) for the $J$th remote node is given by: 
\begin{equation}
\begin{split}
\textbf{s}^{(J)} &= [s^{(J)}_{0,0} \quad M^{(J)}_1\quad M^{(J)}_2 \quad ...\quad M^{(J)}_B]^T,\\
M^{(J)}_k &= [s^{(J)}_{k,0} \quad s^{(J)}_{k,1} \quad  ... \quad s^{(J)}_{k,\Omega-1} ]_{1 \times \Omega} , \ \  k = 1,2, ... , B,
\end{split}
\label{sJ}
\end{equation}
where $s^{(J)}_{i,j}$ for $0\leq i \leq B$ and $0\leq j \leq \Omega-1$ is the probability that the transmit buffer assigned to the $J$th remote node is in state $\alpha_{i,j}$.  
At steady state, the probability $s^{(J)}_{i,j}$ of being in any state $\alpha_{i,j}$ will not change with time. In this case, the distribution vector $\textbf{s}^{(J)}$ settles down to a unique value and satisfies the equation $ \textbf{P}^{(J)} \ \textbf{s}^{(J)}=\textbf{s}^{(J)}$ \cite{kn:book_Fayez}.
This means that the steady state distribution vector $\textbf{s}^{(J)}$ is the eignvector for $\textbf{P}^{(J)}$ corresponding to the eign value that equals to one.
MATLAB and other mathematical packages such as Maple and Mathematica have commands for finding that eigenvector. Having found a numeric answer, we must normalize $\textbf{s}^{(J)}$ to ensure that $\sum \limits_{i=0}^B  \sum \limits_{j=0}^{\Omega-1}  s^{(J)}_{i,j} = 1$.

We will use the equal priority protocol introduced in \cite{kn:multiuser0} as a bench mark for comparison purposes. Using this protocol, the central node will use the common backup RF link to communicate with any one of the remote nodes with failed FSO links on an equal priority basis. Thus, the probability of using the common RF link to send data frame to $J$th remote node with poor FSO link is defined as \cite{kn:multiuser0}:
\begin{equation}
\label{PRFJ_e}
P^{(J)}_{RF} = \bigg(\frac{1-b}{N}\bigg) \left[1-(1-a)^N\right],  \quad 1\le J \le N.
\end{equation}

\section{Non-equal Priority with $P$-Persistence Protocol}
The remote nodes of the network are arranged in an ascending order according to their priorities ($1, 2, ..., N$). Remote node $\#$ $1$ will have the highest priority and remote node $\# N$ will have the lowest priority. When more than one remote node have their corresponding FSO links failed, the central node will allocate the common backup RF link to the highest priority remote node ((i.e., the node with the smallest order number among this group of nodes needing the common backup RF link) with probability $p$. It will be shown in the numerical section that allocating the common backup RF link to a certain remote node with probability $p$ will improve the performance of the network, especially the performance of the remote nodes with lower priority. As a special case of this protocol will be the non-equal priority protocol when putting the persistence probability $p$ equals to 1.

\subsection{Probability of Assigning RF Link to the 1st Remote Node }
The probability of using the common backup RF link by the central node to transmit data frame to remote node $\#$1 when its corresponding FSO link fails is $P^{(1)}_{RF}$, which is defined by:
\begin{equation}
P^{(1)}_{RF} = a (1-b)p,
\end{equation}
where $1-b$ is the probability that the common backup RF link is in good condition.
\subsection{Probability of Assigning RF Link to the 2nd Remote Node }
The probability of using the common RF link by the central node to transmit data frame to remote node $\#$2 when its corresponding FSO link fails is $P^{(2)}_{RF}$, which is defined by:
\begin{equation}
P^{(2)}_{RF} = a (1-b) p x_1,
\end{equation}
where $x_1$ is the probability that the RF link is not assigned to the first remote node, defined as $x_1 = 1 - y_1$.
$y_1$ is the probability that the central node uses the RF link to send data frame to the first remote node which can be defined as: 
\begin{equation}
\label{y1}
\begin{split}
y_1 = s^{(1)}_{0,0} \omega a (1-b) p + a (1-b) p\sum \limits_{i=1}^B s^{(1)}_{i,0}+ \sum \limits_{i=1}^B  \sum \limits_{j=1}^{\Omega-1} s^{(1)}_{i,j},
\end{split}
\end{equation}
where in (\ref{y1}), the first term on the RHS is the probability that a frame arrives to the transmit buffer for the first node and leaves it using the RF link when the transmit buffer is empty. The second term and the third term on the RHS denotes the probability that a frame leaves the transmit buffer using RF link and the buffer is not empty.
\subsection{Probability of Assigning RF Link to the 3rd Remote Node }
The probability of using the common RF link by the central node to transmit data frame to remote node $\#$3 when its corresponding FSO link fails is $P^{(3)}_{RF}$, which is defined by:
\begin{equation}
P^{(3)}_{RF} = a (1-b)p x_1x_2,
\end{equation}
where $x_2$ is the probability that the RF link is not assigned to the second remote node, defined as $x_2 = 1 - y_2$.
$y_2$ is the probability that the central node uses the RF link to send data frame to the second remote node which can be defined as: 
\begin{equation}
\label{y2}
\begin{split}
y_2 = s^{(2)}_{0,0} \omega a (1-b) p + a (1-b) p \sum \limits_{i=1}^B s^{(2)}_{i,0}+ \sum \limits_{i=1}^B  \sum \limits_{j=1}^{\Omega-1} s^{(2)}_{i,j},
\end{split}
\end{equation}
where in (\ref{y2}), the first term on the RHS is the probability that a frame arrives to the transmit buffer for the second node and leaves it using the RF link when the transmit buffer is empty. The second term and the third term on the RHS denotes the probability that a frame leaves the transmit buffer using RF link and the buffer is not empty.
\subsection{Probability of Assigning RF Link to the $J$th Remote Node }
We can generalize the probability of using the common backup RF link by the central node to transmit data frame to $J$th remote node when its corresponding FSO link fails $P^{(J)}_{RF}$, which is defined by:
\begin{equation}
\label{PRFJ_ne}
P^{(J)}_{RF} = a (1-b) p \prod \limits_{k=0}^{J-1} x_k, \quad x_0 = 1, \quad 1\le J \le N,
\end{equation}
where $x_k$ is the probability that the RF link is not assigned to the $k$th remote node, defined as $x_k = 1 - y_k$.
$y_k$ is the probability that the central node uses the RF link to send data frame to the $k$th remote node which can be defined as: 
\begin{equation}
\label{y2}
\begin{split}
y_k = s^{(k)}_{0,0} \omega a (1-b) p + a (1-b) p \sum \limits_{i=1}^B s^{(k)}_{i,0}+ \sum \limits_{i=1}^B  \sum \limits_{j=1}^{\Omega-1} s^{(k)}_{i,j}.
\end{split}
\end{equation}
\section{Performance metrics for the $J$th remote node}
The steady state distribution vector $\textbf{s}^{(J)}$ defined in (\ref{sJ}) will allow us to analyze the various performance metrics for the $J$th remote node as will be explained in the following subsections. 
\subsection{Throughput from Central Node to the $J$th remote node}
We define the throughput $Th^{(J)}$ from central node to the $J$th remote node to be the probability of successfully transmitting a data frame over the FSO or RF link. In this sense, successful data frame transmission occurs in 3 separable cases which are:
\begin{enumerate}
\item{An arrived data frame is transmitted directly over the FSO or the backup RF link, where there are not any frames stored in the transmit buffer, that is assigned to $J$th remote node, waiting for transmission. 
In this case, the throughput will be given by ${Th_1}^{(J)} =  \omega s^{(J)}_{0,0} \left(P_{FSO} + \frac{1}{\Omega} P^{(J)}_{RF}\right)$, where $s^{(J)}_{0,0}$ is the probability that the transmit buffer assigned to the  $J$th remote node is empty, and $P_{FSO} + \frac{1}{\Omega} P^{(J)}_{RF}$ is the probability that there is an available FSO link or RF link for frame transmission. The $1/\Omega$ factor is because transmitting a data frame over the back-up RF link requires $\Omega$ time steps. Or in other words, $1/\Omega$ of the frame is transmitted over the RF link every time step.}
\item{A data frame leaves the transmit buffer when it is not empty. Specifically, the first frame stored in the transmit FIFO buffer will be transmitted over the FSO link or over the backup RF link if it is available.
In this case, the throughput will be given by ${Th_2}^{(J)} =  \left(P_{FSO} + \frac{1}{\Omega} P^{(J)}_{RF}\right)
\sum \limits_{i=1}^B s^{(J)}_{i,0}$, where $s^{(J)}_{i,0}, i=1,...,B$ is the probability that the transmit buffer of $J$th remote node contains $i$ data frames waiting for transmission.
}
\item{A data frame from the transmit buffer assigned to the $J$th remote node is being transmitted over the backup RF link. However, it needs $\Omega$ time steps to be completely transmitted.\\
In this case, the throughput will be given by ${Th_3}^{(J)} = \frac{1}{\Omega} \sum \limits_{i=1}^B  \sum \limits_{j=1}^{\Omega-1} s^{(J)}_{i,j} $, where $s^{(J)}_{i,j}$ for $1\leq i \leq B$ and $1\leq j \leq \Omega-1$ is the probability that the transmit buffer assigned to the $J$th remote node is in state $\alpha_{i,j}$. Here, $i$ represents the number of data frames stored in the buffer and $j$ represents the number of time steps elapsed since the beginning of one data frame transmission over the RF link. 
}
\end{enumerate}
Thus, the throughput $Th^{(J)}$ can be calculated as:
\begin{equation}
\begin{split}
\label{ThJ}
{Th}^{(J)} & = {Th_1}^{(J)} + {Th_2}^{(J)} + {Th_3}^{(J)}\\
&=  \omega s^{(J)}_{0,0} \left(P_{FSO} + \frac{1}{\Omega} P^{(J)}_{RF}\right) + \left(P_{FSO} + \frac{1}{\Omega} P^{(J)}_{RF}\right) X \\
&\sum \limits_{i=1}^B s^{(J)}_{i,0} + \frac{1}{\Omega} \sum \limits_{i=1}^B  \sum \limits_{j=1}^{\Omega-1} s^{(J)}_{i,j},  \quad 1\le J \le N.
\end{split}
\end{equation}
$P^{(J)}_{RF}$ in (\ref{ThJ}) is given either by (\ref{PRFJ_e}) in case of using equal priority protocol or by (\ref{PRFJ_ne}) in case of using non-equal priority with $p$ persistence protocol.

The optimal $p$ which maximizes the aggregated throughput $Th_{total} =  \sum \limits_{J=1}^N {Th}^{(J)}$, denoted by $p^*$, can be
obtained by solving ${\partial Th_{total}}/{\partial p}=0$ for $p$. However, it is difficult to obtain $p^*$ in closed form. As an alternative method, we can treat the problem as one-dimensional optimization problem with uncertainty range of $[0.001, 1]$ and use numerical methods such as Golden-section search method \cite{kn:book_Lu} to search for the optimal $p^*$.

\subsection{Average Buffer Size}
The average buffer size $Q^{(J)}_a$ is the average number of data frames stored in the transmit buffer and waiting for transmission to the $J$th remote node. Thus, $Q^{(J)}_a$ can be calculated as $Q^{(J)}_a = \sum \limits_{i=1}^B  \sum \limits_{j=0}^{\Omega-1} i  \ s^{(J)}_{i,j},  \quad 1\le J \le N$, which represents the weighted sum of the number $i$ data frames stored in the assigned transmit buffer. 

\subsection{Average Buffer Queuing Delay}
The average queuing delay $T^{(J)}_q$ is the average number of time steps that a data frame spends in the transmit buffer before being transmitted. Using Little's result, this queuing delay is given by $T^{(J)}_q = \frac{Q^{(J)}_a}{Th^{(J)}},  \quad 1\le J \le N$ \cite{kn:book_Fayez}.

\subsection{Frame Loss Probability}
A frame is lost when it arrives to a full transmit buffer and the frame at the head of the buffer does not leave. 
Using traffic conservation principle described in \cite{kn:book_Fayez}, the frame loss probability $P^{(J)}_L$ can be given as $P^{(J)}_L = \omega - Th^{(J)},  \quad 1\le J \le N$. 
\subsection{Efficiency of the Queue}
The efficiency $\varphi^{(J)}$  of data frames queue assigned to $J$th remote node is defined as the ratio of probability of a data frame leaving the transmit buffer relative to the probability that a new data frame arriving at the transmit buffer. Thus, the efficiency $\varphi^{(J)}$ can be expressed as $\varphi^{(J)} = \frac{Th^{(J)}}{\omega},  \quad 1\le J \le N$. 
The efficiency $\varphi^{(J)}$ gives an indication of data frame loss due to transmit buffer overflow. A value of $\varphi^{(J)}=1$ implies no buffer overflow. A value of $\varphi^{(J)} < 1$ implies buffer overflow and potential frame loss. 

\subsection{RF Link Utilization}
The probability that the RF link is needed in a given time step, denoted by $N_e$, can be calculated as:
\begin{equation}
\begin{split}
N_e&= 1 - \{{P_r}[\gamma_{FSO} \ge \gamma_T]\}^N  = 1 - (1 - a)^N.
\end{split}
\end{equation}
In the non-equal priority with $p$-persistence servicing protocol, the common backup RF link is allocated to the highest priority node with failed FSO link (among a group of remotes nodes with failed FSO links) with probability $p$. Thus, the RF link utilization $U$ defined as the probability that the RF link is used in a given time step can be calculated as $U = (1-b)p N_e$. 

\section{numerical results}
\begin{figure}[h]
\centering
\includegraphics [height=6cm, width=8cm]{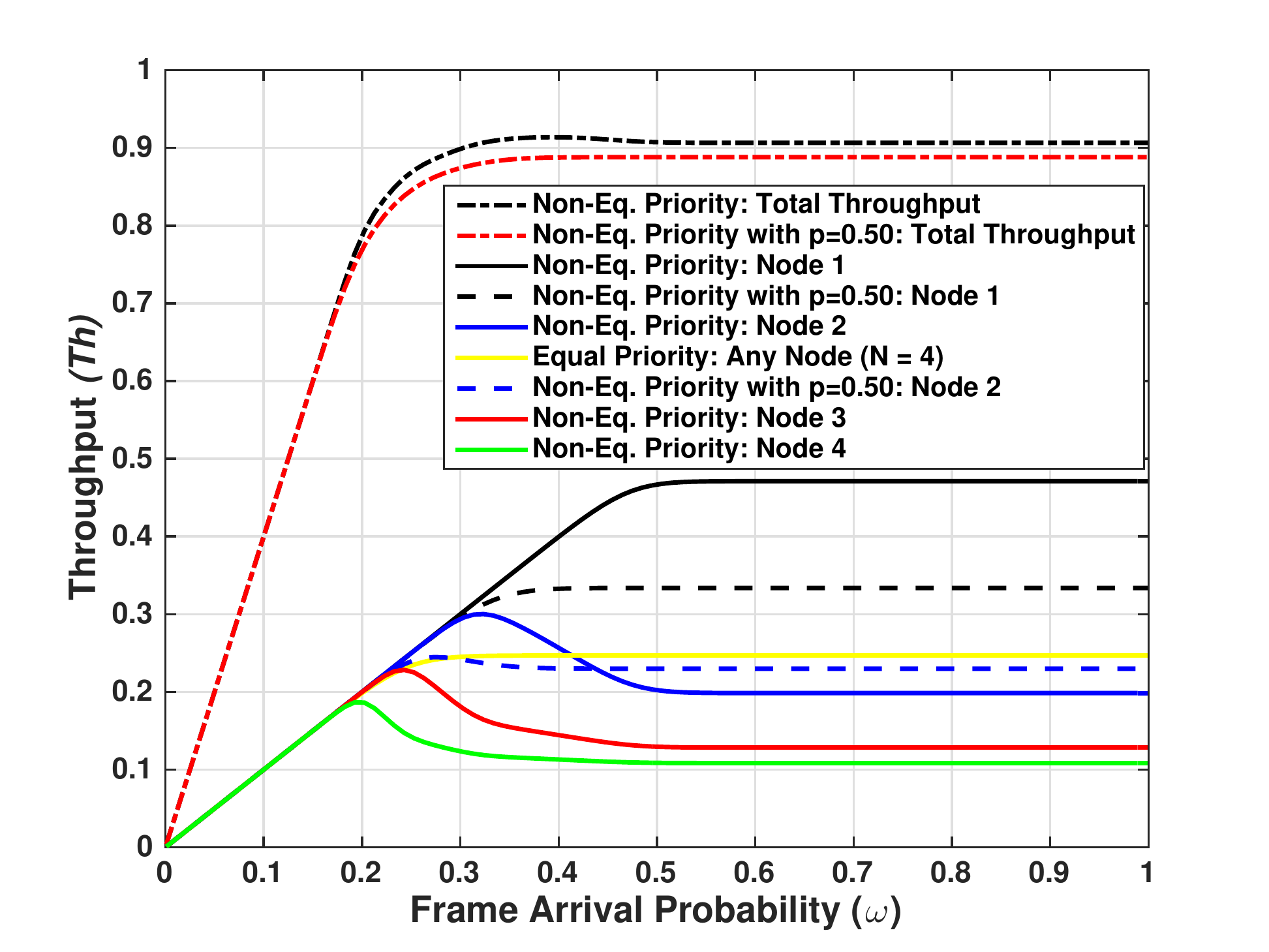} 
\caption{Throughput with $B$ = 10 frames.}
\label{fig:throughput}
\end{figure}

\begin{figure}[h]
\centering
\includegraphics [height=6cm, width=8cm]{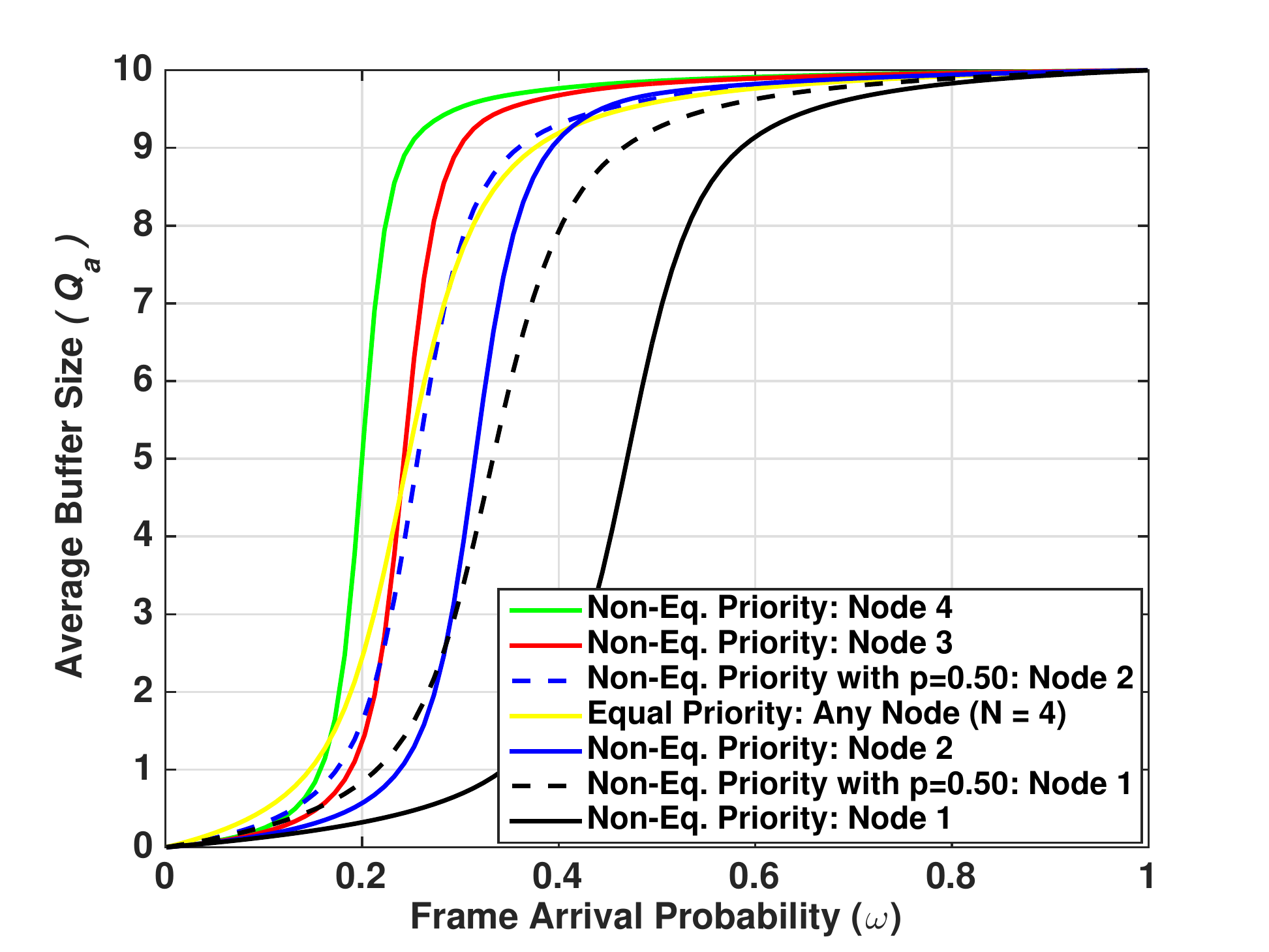} 
\caption{Average buffer size with $B$ = 10 frames.}
\label{fig:Average_Buffer_Size}
\end{figure}

\begin{figure}[h]
\centering
\includegraphics [height=6cm, width=8cm]{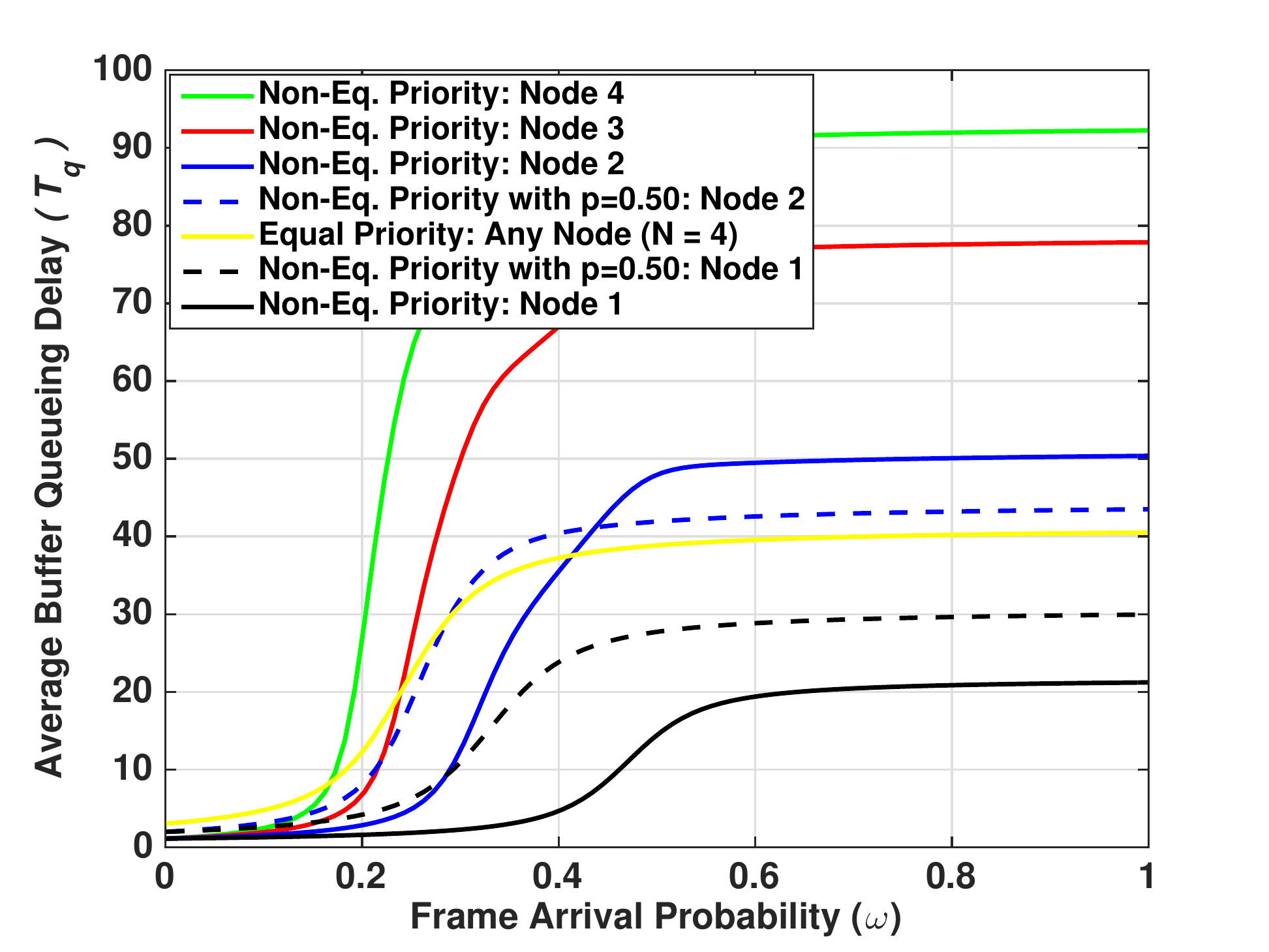} 
\caption{Average queuing delay with $B$ = 10 frames.}
\label{fig:Normalized_avg_queue_delay}
\end{figure}

\begin{figure}[h]
\centering
\includegraphics [height=6cm, width=8cm]{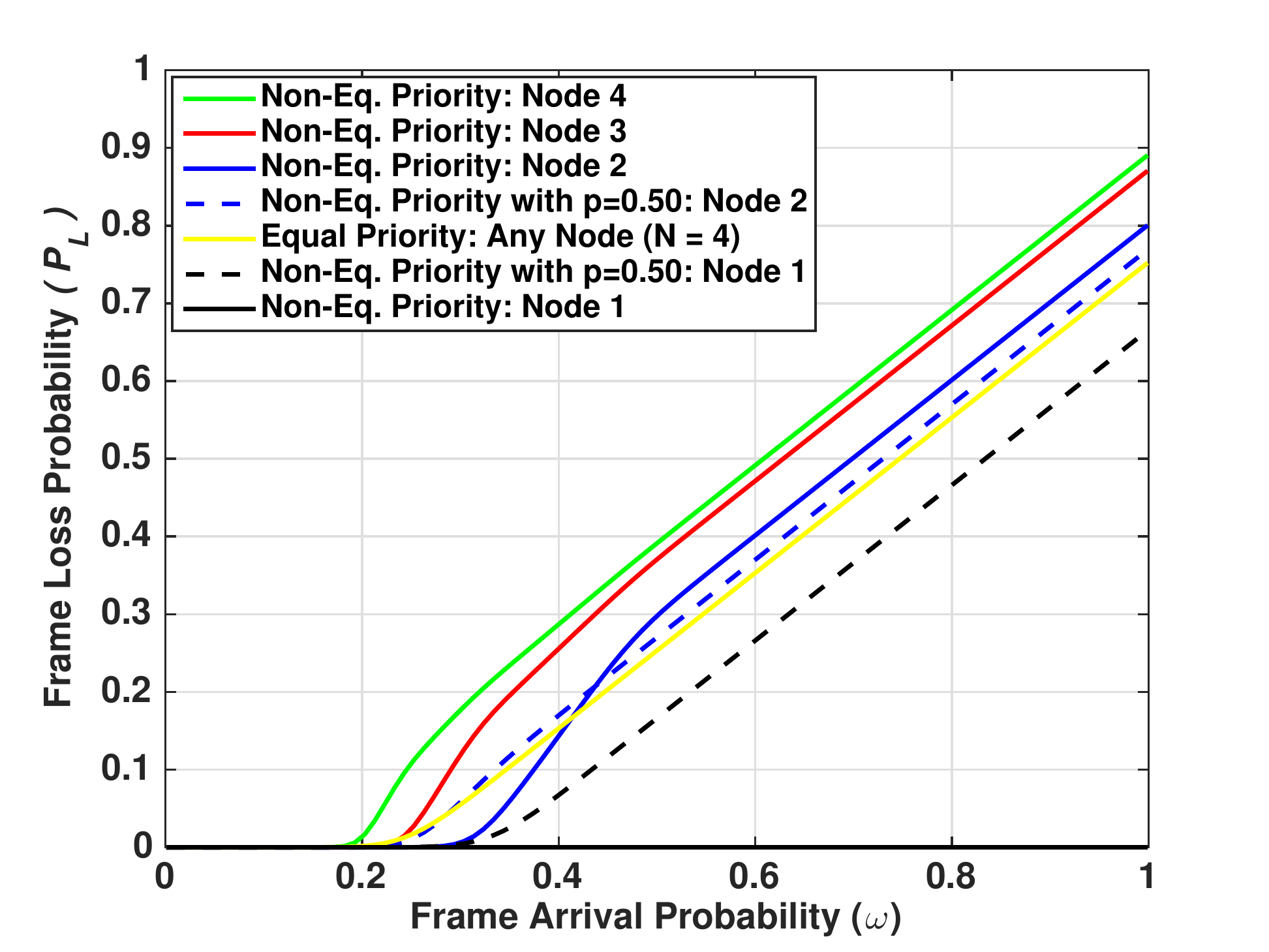} 
\caption{Frame loss probability with $B$ = 10 frames.}
\label{fig:SymbolLoss}
\end{figure}

\begin{figure}[h]
\centering
\includegraphics [height=6cm, width=8cm]{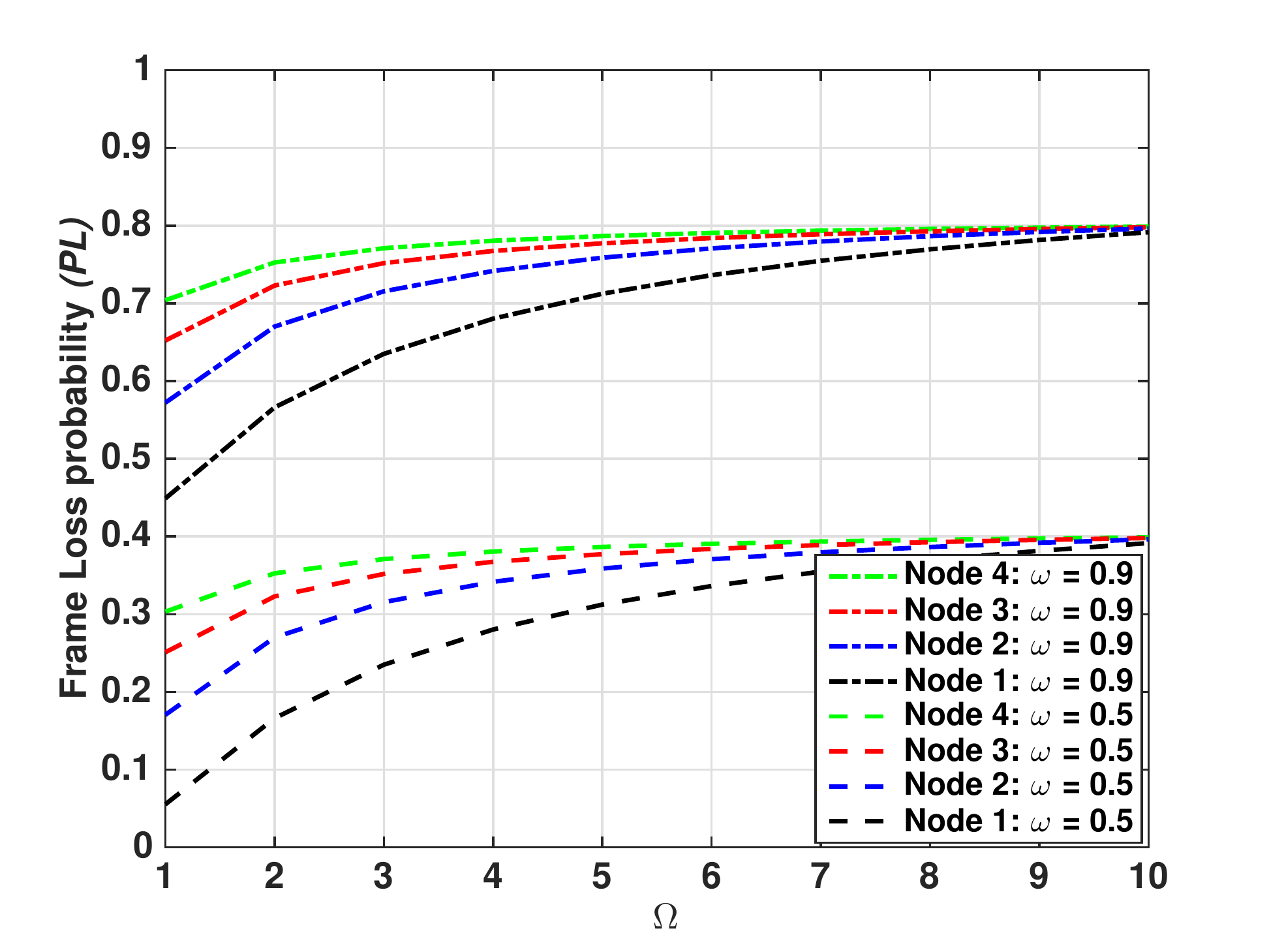} 
\caption{Frame loss probability with $B$ = 10 frames.} 
\label{fig:SymbolLoss_omega}
\end{figure}

\begin{figure}[htp]
\centering
\includegraphics [height=6cm, width=8cm]{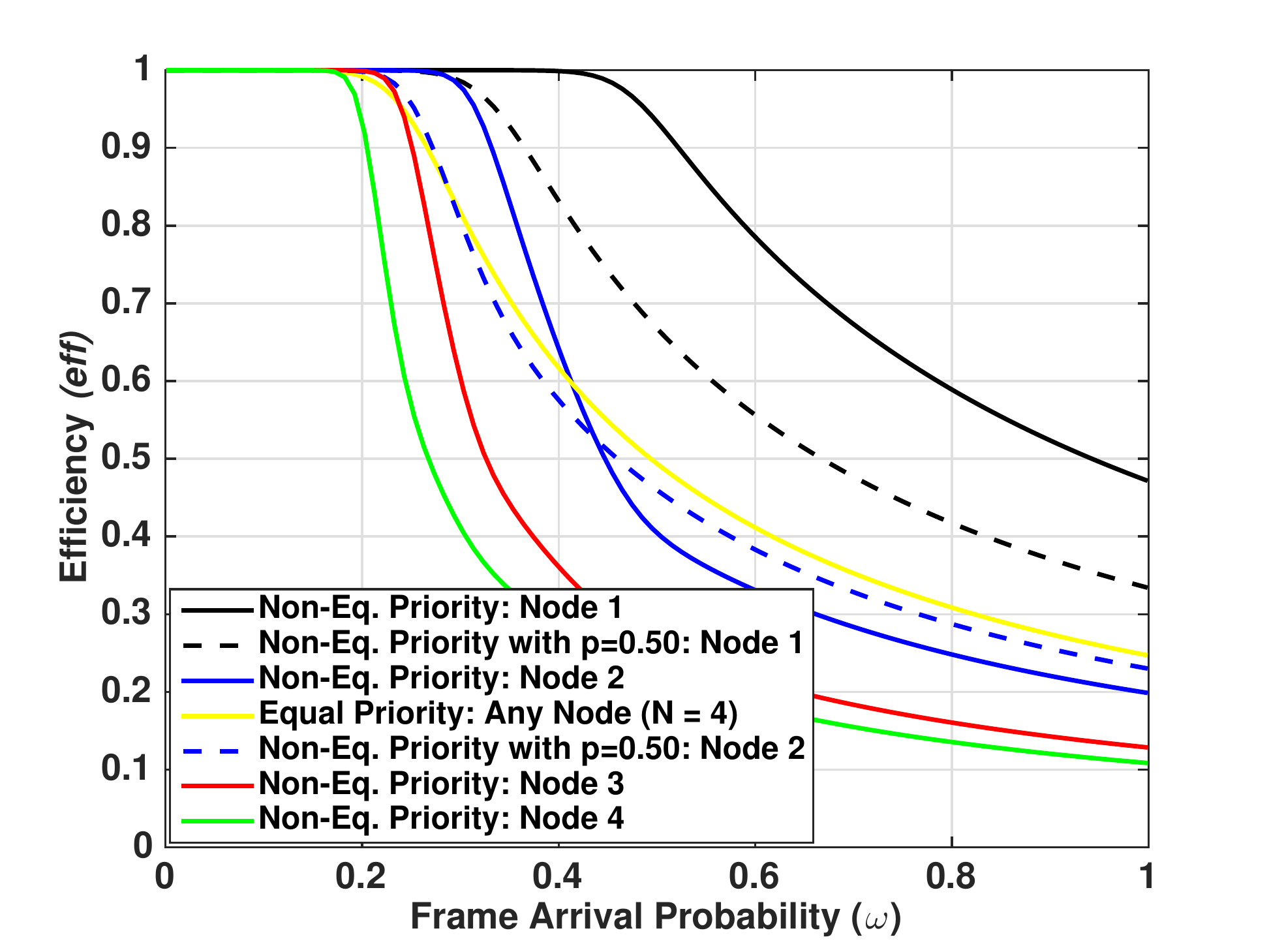} 
\caption{Efficiency with $B$ = 10 frames.}
\label{fig:efficiency}
\end{figure}

\begin{figure}[ht]
\centering
\includegraphics [height=6cm, width=8cm]{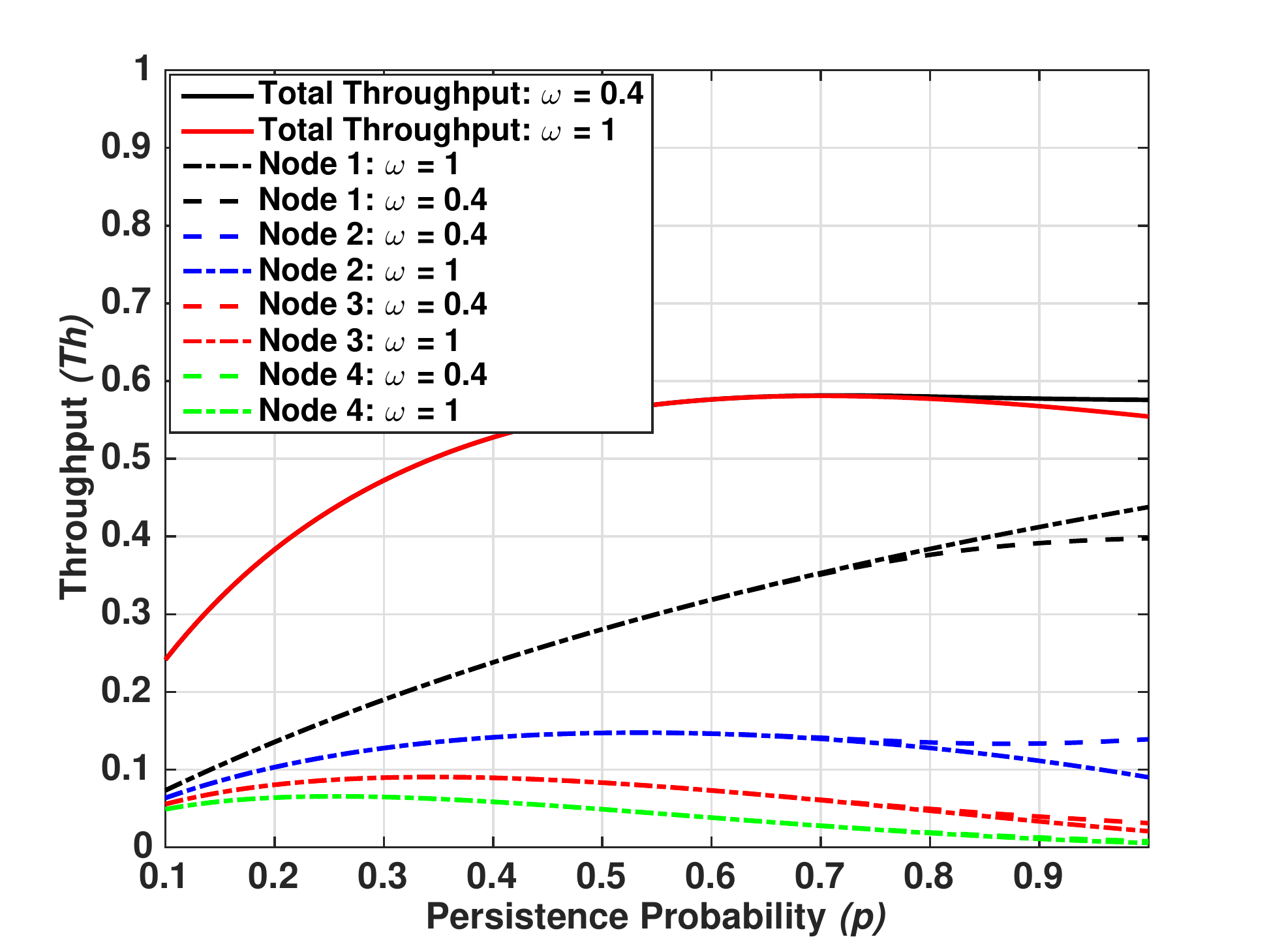} 
\caption{Throughput with $B$ = 10 frames under strong atmospheric turbulence and sever pointing error conditions.}
\label{fig:throughput_p}
\end{figure}

\begin{figure}[h]
\centering
\includegraphics [height=6cm, width=8cm]{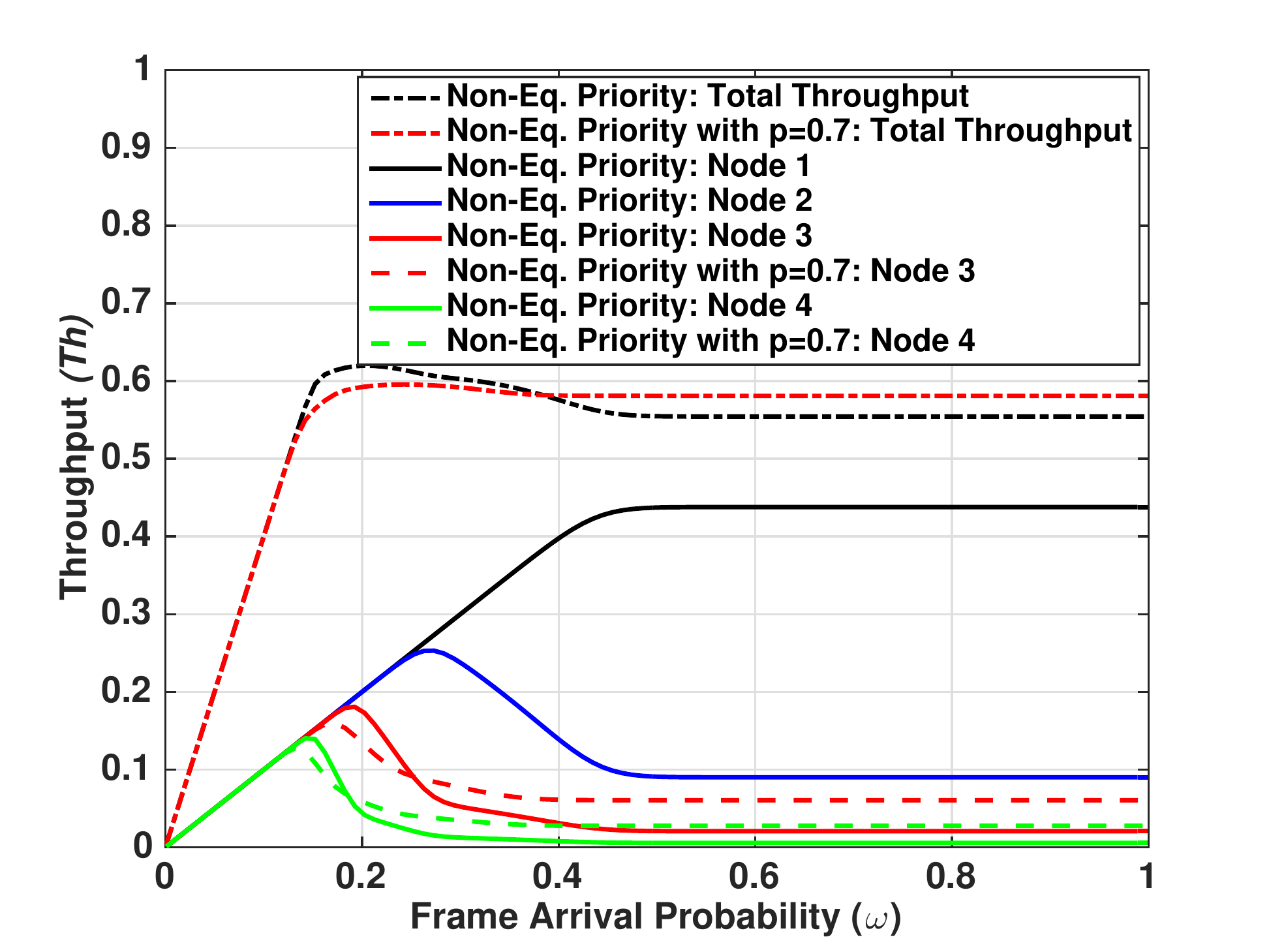} 
\caption{Throughput with $B$ = 10 frames under strong atmospheric turbulence and sever pointing error conditions.}
\label{fig:throughput_strong}
\end{figure}

\begin{figure}[h]
\centering
\includegraphics [height=6cm, width=8cm]{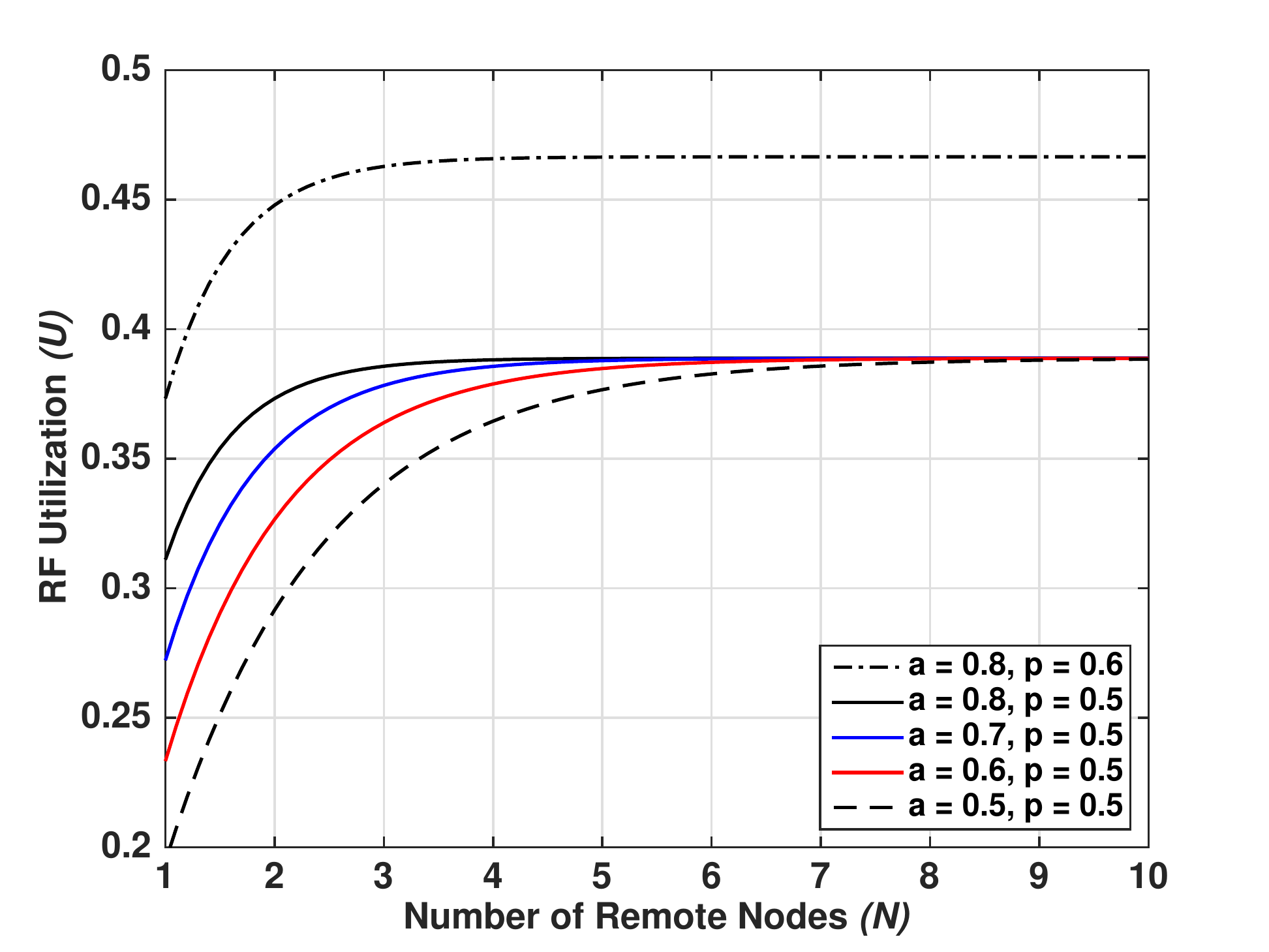}
\caption{RF channel utilization}
\label{fig:RF_utilization}
\end{figure}

In this section, we present several numerical examples to illustrate our analysis. We assume $m$=5, for which Nakagami-$m$ distribution perfectly models fading over MMW RF channel. Assume using 16-QAM digital modulation, $\gamma_T$ is chosen to be equal to 21 dB to satisfy a minimum target BER of $10^{-6}$.  
The relevant parameters of the FSO and RF sub€"systems considered for the numerical results in this paper are provided in Table \ref{table:system_parameter} \cite{kn:Farid}, \cite{kn:ref10}, and \cite{kn:julian2}. These parameters are assumed to be the same for the $N$ remote nodes.
The average QAM symbol energy $E_{avg}$ is assumed to be normalized to unity. 
Mainly, the FSO link is affected by foggy weather. Thus,
we consider in our numerical examples the scenario of moderate foggy weather condition with weather attenuation coefficient $\alpha_{FSO}$ = 42.2 dB/Km and moderate atmospheric turbulence with $C_n^2=5 \times 5^{-14}$ and no rain with RF rain attenuation $\alpha_{rain}$ = 0 dB/Km. Note that, under foggy weather condition and no rain and by using the parameters given in Table \ref{table:system_parameter}, the probabilities $a$ and $b$ are calculated to be equal to $a = 0.90$ and $b=0.22$. These values of $a$ and $b$ are used in Figs. \ref{fig:throughput} - \ref{fig:efficiency}. We use an arbitrary value of $p$ = 0.5 in Figs. \ref{fig:throughput} - \ref{fig:efficiency}, to investigate the network performance under the proposed non-equal priority with $p$-persistence servicing protocol as opposed to the non-equal priority servicing protocol. The value of $p$ is controlled and optimized to improve the network performance as is shown in Figs. \ref{fig:throughput_p} and \ref{fig:throughput_strong}.
Also, we investigate the network performance under strong atmospheric turbulence ($\alpha$ = 2.064 and $\beta$ = 1.342) and sever pointing error ($\xi$ = 1.1) \cite{kn:strong} conditions in Figs. \ref{fig:throughput_p} and \ref{fig:throughput_strong}. In Fig. \ref{fig:RF_utilization} we use arbitrary values of $a$ with $b=0.22$. We assume that the frame transmission rate of FSO link is twice that of the RF link (i.e., $\Omega = 2$) in plotting all Figures expect Fig. \ref{fig:SymbolLoss_omega}, where it is plotted against range of $\Omega$'s values.
The analytical results  have been verified by using Monte-Carlo simulation.
\begin{table}[h] 
\caption{Parameters of FSO and RF subsystems} 
\centering 
\small\addtolength{\tabcolsep}{0.4pt}
\begin{tabular}{||c||c||c||}
  \hline\hline
  Parameter & Symbol & Value\\
  \hline\hline 
  \multicolumn{3}{||c||}{FSO Subsystem}\\
  \hline\hline
  Wavelength  & $\lambda_{FSO}$ & 1550 nm  \\
  Oscillator Power & $P_{LO}$ & $10^{-2}$ W\\
 Shot Noise Variance & $\sigma_{FSO}^2$ & $5 \times 10^{-12}$ \\
  Responsivity    & $\eta$ & 0.5 A/W   \\
  Photodetector Diameter    & D & 20 cm  \\
  Transmit Power & $P_{FSO}$ & 15 dBm \\
  Transmit divergence at $1/e^2$   & $\theta_0$ & 2.5 mrad  \\
  Jitter standard deviation  & $\sigma_s$ & 30 cm  \\
  Link distance & $z$ & 1000 m  \\
  \hline\hline
  \multicolumn{3}{||c||}{RF Subsystem}\\
  \hline\hline
  Carrier Frequency  & $f_{RF}$ & 60 GHz    \\
  Bandwidth & $W$ & 250 MHz    \\
 Transmit Power & $P_{RF}$ & 25 dBm    \\
 Transmit Antenna Gain& $G_T$ & 43 dBi    \\
  Receive Antenna Gain & $G_R$ & 43 dBi    \\
Noise Power Spectral Density & $N_0$ & -114 dBm/MHz    \\
Receiver Noise Figure & $N_F$ & 5 dB    \\
Oxygen Attenuation& $\alpha_{oxy}$ & 15.1 dB/Km    \\
\hline\hline
\end{tabular}
\label{table:system_parameter} 
\end{table}

As an overall observation, using non-equal priority protocol with $p$-persistent probability less than 1 improves the performance of the remote nodes which approaches the performance when using the equal priority protocol. 

In Fig. \ref{fig:throughput}, we plot the throughput $Th$ as function of the frame arrival probability $\omega$. Considering the non-equal priority protocol ($p$ = 1), it can be seen that the throughput goes into three phases viz linear, dip and saturation. It is linear for small values of $\omega$. The linearity is because a frame arrives and leaves the buffer immediately. This means that the buffer remains empty in this linear phase.  The saturation phase occurs for higher values of $\omega$. For node 1, there is no dip phase and the transition from linear to saturation phase occurs when the transmit buffer is not empty. Nodes 2, 3 and 4 experience sudden drops in throughput after the linear phase; the transition occurs when the transmit buffer is not empty and arriving frames can not leave the buffer because access to the common RF link is restricted. As we have used a non-equal priority protocol ($p$ = 1), node 1 has highest priority to access the RF link and priority decreases with increase in node number. The access to the common RF link for a node is restricted by a buffer of lower number node, which causes the drop in throughput for nodes 2, 3 and 4. 

Also, it can be seen from Fig. \ref{fig:throughput} that the throughput graph, when using non-equal priority protocol with $p$-persistent, experiences the same behaviour as the case when using non-equal priority protocol ($p$ = 1). However, the throughput performance of the nodes 2, 3, and 4 is improved (node 2 is only shown in Fig. \ref{fig:throughput}). The $p$-persistent value is considered as $p$ = 0.50, which means that even when the common RF link is available, the probability of accessing it for every remote node is only 50$\%$. At the same time, throughput performance for node 1 decreases as its chance to access the common RF link is lowered to 50$\%$.  
Also, it can be seen that the total throughput of the network is improved when using non-equal priority with $p$-persistence servicing protocol, which approaches the performance when using non-equal priority servicing protocol.

In Fig. \ref{fig:Average_Buffer_Size}, we plot the average buffer size $Q_a$ as function of the frame arrival probability $\omega$.
Considering the non-equal priority protocol ($p$ = 1), it can be seen that for the same value of frame arrival probability $\omega$, the transmit buffer assigned for remote node 4 has the largest number of frames waiting for transmission. This is because it has the lowest priority of accessing the common RF link in case of the failure of its corresponding FSO link. As the order of the remote node decreases, its corresponding assigned transmit buffer contains less number of waiting frames. This is because its priority of accessing the common RF link increases.
The average buffer size $Q_a$ saturates at the maximum buffer size $B$ as expected.

Also, it can be seen from Fig. \ref{fig:Average_Buffer_Size} that the average buffer size graph, when using non-equal priority protocol with $p$-persistent, experiences the same behaviour as the case when using non-equal priority protocol ($p$ = 1). However, the average buffer size for the nodes 2, 3, and 4 is improved (node 2 is only shown in Fig. \ref{fig:Average_Buffer_Size}). The $p$-persistent value is considered as $p$ = 0.50, which means that even when the common RF link is available, the probability of accessing it for every remote node is only 50$\%$. This will affect in return the frame waiting time in the corresponding transmit buffer and thus the average number of frames stored in the transmit buffer waiting for transmission. 

In Fig. \ref{fig:Normalized_avg_queue_delay}, we plot the average queuing delay $T_q$ as function of the frame arrival probability $\omega$. The average queuing delay $T_q$ shows an S-type behaviour, where it starts at low values then starts increasing after a certain frame arrival probability value. It then saturates. Considering non-equal priority protocol ($p$ = 1), the saturation value increases with increasing the order of the remote node where its priority of accessing the common RF link decreases. It is noted that the average queuing delay $T_q$ is not equal to zero at very low value of $\omega$. This is because, a frame will wait in the buffer until a link is available for its transmission. 

Also, it can be seen from Fig. \ref{fig:Normalized_avg_queue_delay} that the average queuing delay $T_q$ graph, when using non-equal priority protocol with $p$-persistent, experiences the same behaviour as the case when using non-equal priority protocol ($p$ = 1). However, the average buffer size for the nodes 2, 3, and 4 is improved (node 2 is only shown in Fig. \ref{fig:Normalized_avg_queue_delay}). The $p$-persistent value is considered as $p$ = 0.50, which means that even when the common RF link is available, the probability of accessing it for every remote node is only 50$\%$. This will affect in return the frame waiting time in the corresponding transmit buffer. 

In Fig. \ref{fig:SymbolLoss}, we plot the frame loss probability $P_L$ as function of the frame arrival probability $\omega$. The frame loss probability goes into two phases viz zero loss and linear. The zero loss phase occurs at low $\omega$ values. The linear phase occurs for higher values of $\omega$. The transition point occurs when the transmit buffer is full. Considering non-equal priority protocol ($p$ = 1), it can be seen that node 1 shows zero loss. This is because Node 1 has the highest priority to access the common RF link and to transmit the frames in its corresponding transmit buffer, hence the transmit buffer never gets full. The nodes 2, 3 and 4 transition points occur at lower values of $\omega$. This is because the common RF link access priority decreases with increase in node number and the corresponding transmit buffer starts filling up at lower value of $\omega$.

Also, it can be seen from Fig. \ref{fig:SymbolLoss} that the frame loss probability $P_L$ graph, when using non-equal priority protocol with $p$-persistent, experiences the same behaviour as the case when using non-equal priority protocol ($p$ = 1). However, the frame loss probability for the remote nodes 2, 3, and 4 is improved (node 2 is only shown in Fig. \ref{fig:SymbolLoss}) due to increasing the remote nodes chance of accessing the common RF link when their corresponding FSO links fails. The $p$-persistent value is considered as $p$ = 0.50, which means that even when the common RF link is available, the probability of accessing it for every remote node is only 50$\%$. This will lower in return the frame waiting time in the corresponding transmit buffer and thus decreases the frame  loss probability. It can be seen from Fig. \ref{fig:SymbolLoss} that the frame loss probability $P_L$ when using non-equal priority protocol with $p$-persistent for node 1 experiences linear behaviour because its chance to access the common RF link is lowered to 50$\%$. when using the equal priority protocol.

In Fig. \ref{fig:SymbolLoss_omega}, we plot the frame loss probability $P_L$ as function of $\Omega$ for certain value of $p$ = 0.5. At certain value of frame arrival probability $\omega$, the frame loss probability increases as $\Omega$ increases. This is because transmitting a frame over the backup RF link requires more time steps as $\Omega$ increases. This means that more frames will accumulate in the transmit buffers, which in return will increase the probability of loss of new arriving frames. As $\omega$ increases, the frame loss probability increases because more frames arrive at a full transmit buffer while the backup RF link is still busy in transmitting the current frame.

In Fig. \ref{fig:efficiency}, we plot the efficiency $\varphi$ as function of the symbol arrival probability $\omega$.  
The efficiency $\varphi$ goes into two phases viz unity efficiency and decreasing efficiency. The transition points in efficiency occur when the corresponding transmit buffer is not empty which is noted to be happen at lower values of $\omega$. This is because the access to the common RF link for a remote node is restricted by a node of lower number and thus the arriving frames are accumulated in the transmit buffers assigned for the other remote nodes that have failed FSO links.
Also, it can be seen from Fig. \ref{fig:efficiency} that the efficiency $\varphi$ graph when using non-equal priority protocol with $p$-persistent, experiences the same behaviour as the case when using non-equal priority protocol. However, the efficiency for the remote nodes 2, 3, and 4 is improved (node 2 is only shown in Fig. \ref{fig:efficiency}). This is because of persistent probability, which transmits frame with probability $p$, increasing the chances for a frame being transmitted to all the remote nodes. 

In Fig. \ref{fig:throughput_p}, we plot the throughput $Th$ as function of the persistence probability $p$. 
We consider in Fig. \ref{fig:throughput_p} two different values of frame arrival probability, a low value of $\omega$ = 0.4 and high value of $\omega$ = 1. It can be seen that the throughput of node 1 increases as $p$ increases. This is because it has the highest priority for being serviced by the common RF link when its FSO link fails. While, the throughputs of the lower priority nodes 2, 3, and 4 increase and then decrease as $p$ increases. By using Golden-section search method, the optimum $p$ that maximizes the total throughput was found to be $p^* = 0.7$ in this case. It is clear that lower values of $\omega$ gives better throughput performance. This is because less number of frames are arrived at the transmit buffers of the nodes and thus the probability of successful frames transmissions increases.

In Fig. \ref{fig:throughput_strong}, we plot the throughput $Th$ as function of the frame arrival probability $\omega$. 
It can be seen that the throughput performance is greatly degraded under these sever turbulences and pointing errors conditions, especially the lower priority nodes' throughputs. The throughput performance gets its worst as the frame arrival probability $\omega$ increases over nearly 0.3. This is because there are new frames arriving at the transmit buffers of nodes 3 and 4, and there are not available link for frame transmissions. Nodes 3 and 4's FSO links are failed and the common backup RF link is being used by higher priority nodes. However, when using the non-equal priority with $p$-persistence servicing protocol and $p^*$ = 0.7, the throughput performance of nodes 3 and 4 is improved. At the same time, the total throughput of the network is also improved, especially for values of $\omega$ greater than 0.4. 

\begin{figure}[h]
\centering
\includegraphics [height=6cm, width=8cm]{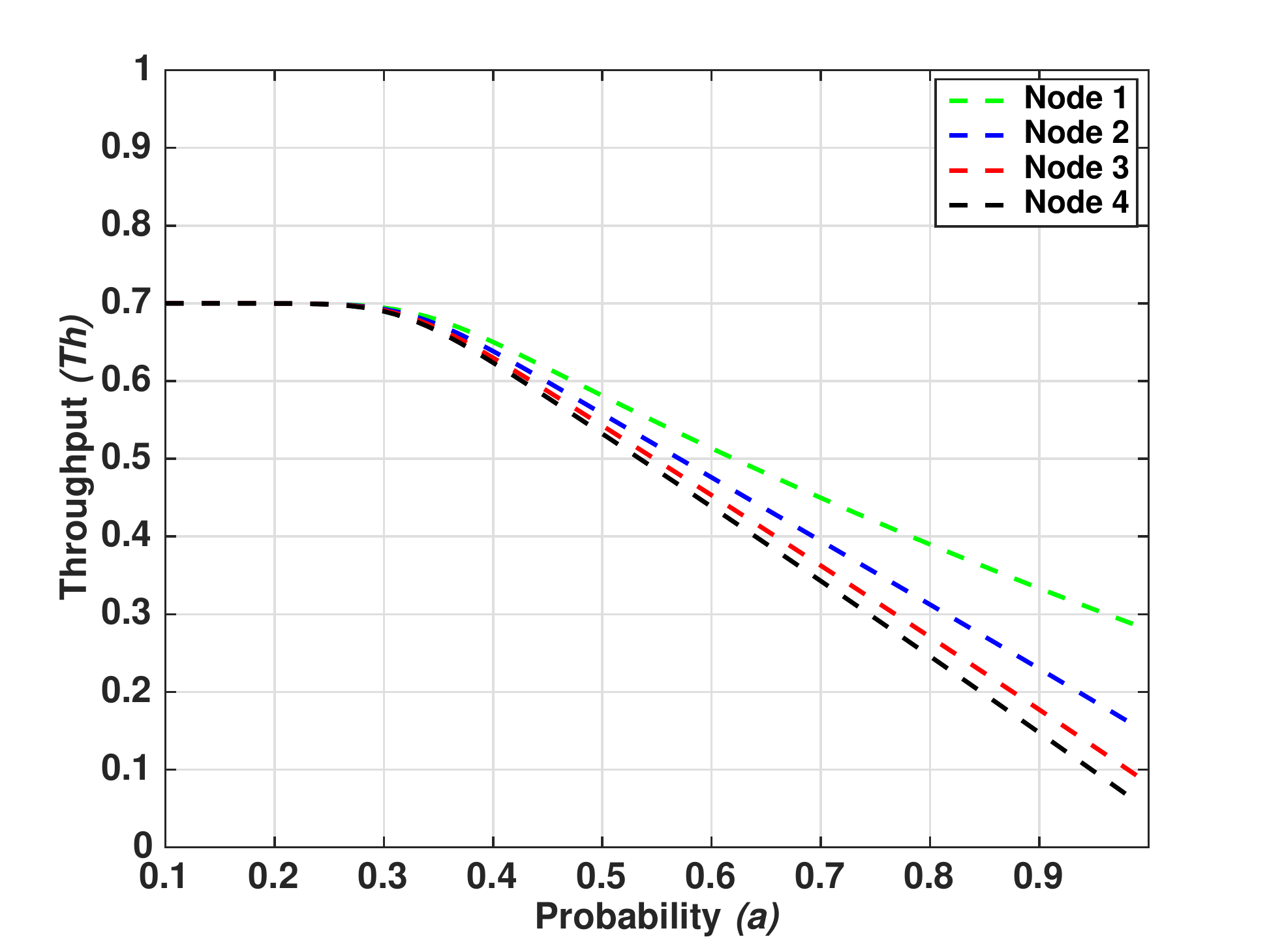} 
\caption{Throughput with $B$ = 10, $p$ = 0.5, $b$ = 0.22, and $\omega$ = 0.7 .}
\label{fig:throughput_a}
\end{figure}

In Fig. \ref{fig:throughput_a}, we plot the throughput $Th$ as function of the probability $a$. It can be seen that for small values of $a$, which means that most of the time the FSO channels are in good conditions, the throughput equals $\omega$ = 0.7. This is due to the successful transmission of arrived frames over the separable FSO links. As the quality of the FSO channels degrades, i.e., $a$ increases, the throughput decreases due to the increased probability of frames loss. As expected, node 1 has the best throughput as it has the highest priority to be serviced by the common RF link when its FSO link fails. However, non-equal priority with $p$-persistence servicing protocol gives the lower priority nodes a chance to receive their intended data when the common backup RF link is in good condition.

\begin{figure}[h]
\centering
\includegraphics [height=6cm, width=8cm]{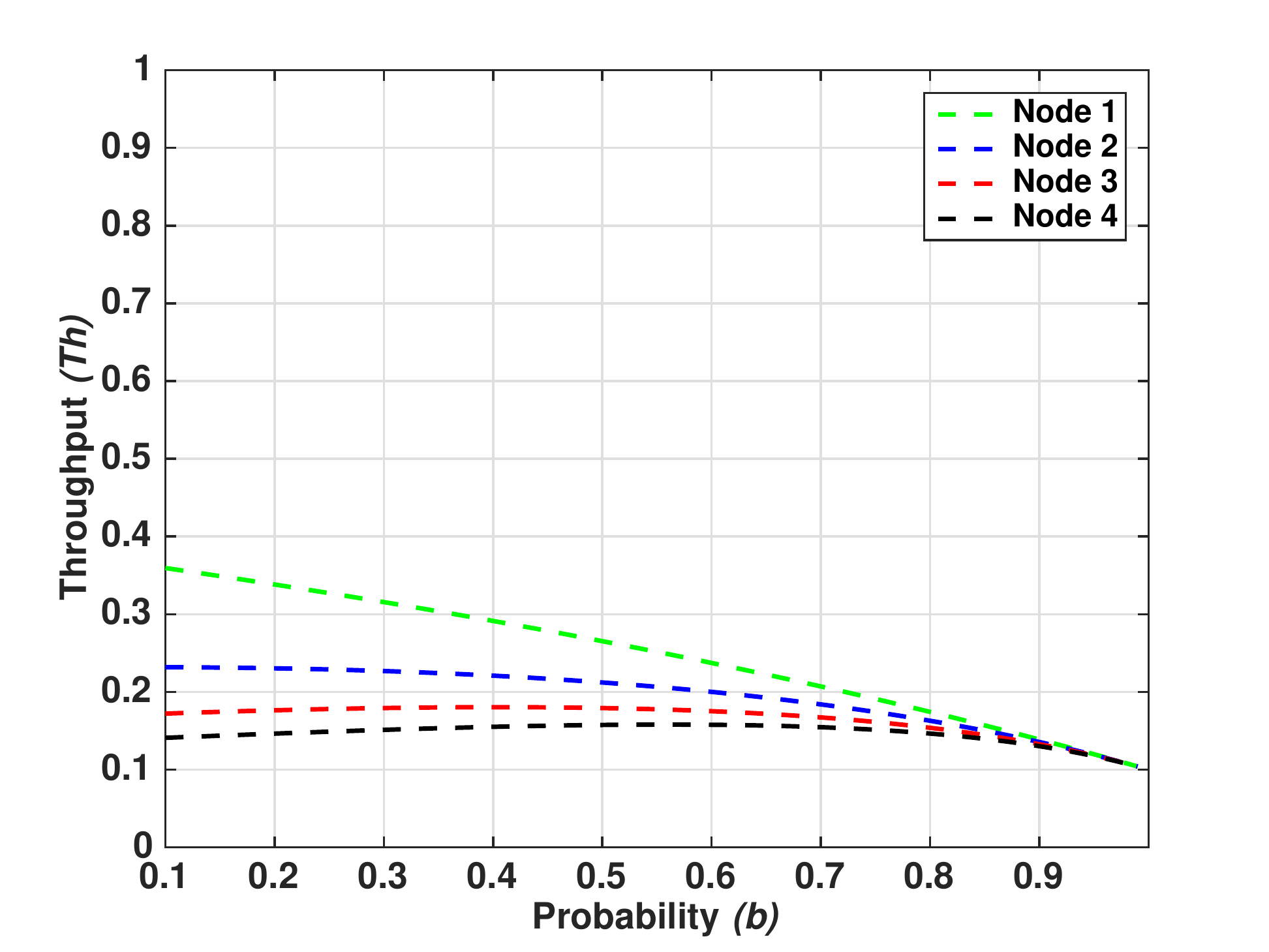} 
\caption{Throughput with $B$ = 10, $p$ = 0.5, $a$ = 0.9, and $\omega$ = 0.7 .}
\label{fig:throughput_b}
\end{figure}
In Fig. \ref{fig:throughput_b}, we plot the throughput $Th$ as function of the probability $b$. We choose  $a$ = 0.9 to indicate that the FSO links are in poor condition. It can be seen that the throughput performance degrades as $b$ increases. This is due to the degradation of the quality of the common backup RF link and thus the probability of frames loss increases. As expected, node 1 has the best throughput as it has the highest priority to be serviced by the common RF link when its FSO link fails. While, the lower priority nodes 2, 3, and 4 have lower throughput performance. 

In Fig. \ref{fig:RF_utilization}, we plot the RF utilization $U$ as function of the number of the remote nodes $N$ for different values of the poor quality probability of the FSO link $a$. The RF utilization $U$ increases as $N$ increases until it saturates. This is because, as $N$ increases the probability that more FSO links are in poor quality increases and thus more remote nodes will need to access the RF link. The value of the saturation depends on the quality of the RF link and the value of $p$.
For a fixed value of $N$, the RF link utilization increases as the FSO link quality degrades. For a fixed value of $N$ and under certain condition of FSO link, the RF link utilization increases as the value of $p$ increases (i.e, as the probability of allocating the common backup RF link to a certain remote user increases).

\section{CONCLUSION} 
In this paper, we proposed a novel multiuser network based on hybrid FSO/RF transmission system. It was assumed that servicing different remote nodes in the network follows a non-equal priority protocol where the user priority decreases as its order in the network increases.
A common backup RF link is used by the central node for data transmission to any remote node in case of the failure of its corresponding FSO link. A discrete-time Markov chain model was developed for the transmit buffer assigned for each remote node where different transmission rates over both RF and FSO links is assumed.
we investigated several performance criteria such as throughput from central node to the remote node, the average transmit buffer size, the frame queuing delay in the transmit buffer, the efficiency of the queuing system, the frame loss probability, and the RF link utilization.
Remote node 1 had the best performance among the other remote nodes as it had the highest priority. As the order of the remote nodes increased the corresponding performance got worse. It was found that transmitting data frames over the common backup RF link with probability $p$ had improved the performance of all the remote nodes and thus the performance of the overall network.
The performance improvement had approached the performance of the network when using equal priority protocol to serve all the remote nodes. The non-equal priority with $p$-persistence servicing protocol provided a compromise to give better performance to high-priority nodes while allowing lower priority nodes a chance to receive their intended data.
As a future work, more than one backup RF channels can be used to serve the nodes with failed FSO links. This will lead to a new Markov chain model for the multiuser network and new system performance analysis.

\end{document}